\documentclass[11pt]{article}
\usepackage[utf8]{inputenc}
\usepackage{amsmath}
\usepackage{bm}
\usepackage{cancel}  
\usepackage{amssymb}
\usepackage{amsthm}
\usepackage[titletoc,toc,title]{appendix}
\usepackage{array}
\usepackage{authblk}
\usepackage{bbm}
\usepackage{caption}
\usepackage{underscore}
\usepackage{dsfont}
\usepackage{setspace}
\usepackage{enumitem}
\usepackage[margin=1in]{geometry}
\usepackage{graphicx}
\usepackage{hyperref}
\usepackage{layout}
\usepackage{mathtools}
\usepackage{multirow}
\usepackage{afterpage}
\usepackage[table,xcdraw]{xcolor}
\usepackage{url}
\usepackage[
backend=biber, natbib,
citestyle = authoryear,
style = authoryear,
uniquelist=false,
uniquename = false,
sorting = nyt,
maxbibnames=99,
maxcitenames=2,
]{biblatex}
\renewbibmacro{in:}{}
\usepackage{leftidx}
\usepackage{subcaption}
\usepackage{tikz}
\usepackage[ruled,linesnumbered]{algorithm2e}
\usepackage{booktabs}
\usepackage{lscape}
\usepackage{mathrsfs}
\usepackage[T1]{fontenc}

\usetikzlibrary{shapes, shadows, arrows, positioning, decorations.markings, decorations.pathreplacing, matrix, calc}

\usetikzlibrary{arrows.meta}

\makeatletter
\providecommand\phantomcaption{\caption@refstepcounter\@captype}
\makeatother

\title{Entity-Specific Cyber Risk Assessment Using Insurtech-Empowered Risk Factors}

\author[$\dagger$]{Jiayi Guo}
\author[$\star$]{Linfeng Zhang}
\author[$\dagger$]{Zhiyu Quan}

\affil[$\dagger$]{Actuarial and Risk Management Sciences, University of Illinois at Urbana-Champaign. Email: jg54@illinois.edu.}
\affil[$\star$]{Department of Mathematics, The Ohio State University. Email: zhang.14673@osu.edu.}
\affil[$\dagger$]{Actuarial and Risk Management Sciences, University of Illinois at Urbana-Champaign. Email: zquan@illinois.edu.}
\begin{document}

%\sloppy

\theoremstyle{definition}
\newtheorem{theorem}{Theorem}[section]
\newtheorem{corollary}[theorem]{Corollary}
\newtheorem{lemma}[theorem]{Lemma}
\newtheorem{proposition}[theorem]{Proposition}

\newtheorem{definition}{Definition}[section]
\newtheorem{problem}{Problem}
\newtheorem{remark}{Remark}[section]
\newtheorem{example}{Example}[section]
\setcounter{section}{0}
\date{}
\maketitle

\begin{abstract}

The lack of high-quality public cyber incident data limits empirical research and predictive modeling for cyber risk assessment. This challenge persists because companies are reluctant to disclose incidents that could damage their reputation or investor confidence. From an actuarial perspective, potential resolutions include these: the enhancement of existing cyber incident datasets and the implementation of advanced modeling techniques to optimize the use of the available data. A review of existing data-driven methods highlights a significant lack of entity-specific organizational features in publicly available datasets. To address that gap, we propose a novel insurtech framework that enriches cyber incident data with entity-specific organizational features. We develop various machine learning (ML) models: a multilabel classification model to understand the occurrence of cyber incident types (e.g., privacy violation, data breach, fraud and extortion, IT error, and others) and a multi-output regression model to estimate their annual frequencies. While classifier and regressor chains are also implemented to explore dependencies among cyber incident types, no significant correlations are observed in our datasets. We also apply multiple interpretable ML techniques to identify and cross-validate potential risk factors developed by insurtech across ML models. We find that compared with conventional risk factors, insurtech-empowered features enhance occurrence and frequency estimation robustness. The framework generates transparent, entity-specific cyber risk profiles, supporting customized underwriting and proactive cyber risk mitigation. It provides insurers and organizations with data-driven insights to support decision-making and compliance planning.

\end{abstract}

\section{Introduction}\label{sec:intro}

Cyber incidents, such as data breaches and ransomware attacks, pose significant threats to modern businesses. The consequences of operational cyber risk can compromise the confidentiality, availability, or integrity of information or information systems and ultimately result in financial loss (See \citet{cebula_taxonomy_2014}). To mitigate such impacts, many organizations have turned to cyber insurance, which has become an integral component of enterprise risk management. From the insurance industry's perspective, cyber risk assessment is a crucial component in underwriting practices, and it typically relies on historical experiences with incidents that caused material losses. However, as identified in \citet{biener_insurability_2015}, the scarcity of data on historical cyber incidents is a major barrier that hinders the development of the cyber insurance market because it prevents pricing models from being effectively evaluated. More recent studies, such as that of \citet{zangerle_modelling_2023}, suggest that although marginal improvements have been made over the years, the shortage of data remains an issue. As well as insurers, organizations in need of a better understanding of their own cyber risks also have difficulties in learning from the past to support the assessment. Moreover, the lack of quality data in the public domain limits the ability of researchers to generate insights from empirical studies and build high-performing predictive models.

\subsection{Data scarcity issue}

The scarcity of data is unlikely to be addressed in the short term. To explain why, \citet{romanosky_examining_2016} describes the typical data-generating process used to create existing public or proprietary cyber incident databases. That process consists of a sequence of three actions: the detection, reporting, and recording of a cyber event after the event occurs. Cyber incident detection is a technically challenging task because the consequences of many incidents, such as data breaches, are not immediately tangible. A survey by \citet{ibm_cost_nodate} finds that the average time to detect a data breach is more than 200 days. The reporting of cyber incidents is largely a regulatory problem because companies lack the incentive to report incidents, as such information may very likely deliver a negative signal to customers and investors. To highlight regulators' role in improving cyber incident data, \citet{kesan_empirical_2021} note that the number of recorded data breaches in the US grew substantially after states started to enact data breach notification laws. Incident detection and reporting determine the number of disclosed cyber events that are available to be recorded, and as long as the technical and legal challenges noted persist, an ample amount of cyber incident data will be unavailable. Thus, under the circumstances, the questions that we may be able to address from an actuarial perspective include how we can enrich the existing cyber incident data and how we can apply advanced modeling techniques to the existing data effectively for cyber risk assessment. In the following, we offer an overview of the proposed data-driven approaches in this regard in the literature.

\subsection{Data-driven approaches to cyber risk modeling}

Several cyber incident datasets are available in the public domain, and many studies leverage them for cyber risk modeling with a focus on the frequency and severity of various types of incidents and the dependence among them. For example, \citet{wheatley_addressing_2021} use information on data breach events from the Privacy Rights Clearinghouse (PRC) and the Open Security Foundation to statistically model the frequency and severity of major breaches with different causes. Also based on the PRC data breach dataset, \citet{eling_copula_2018} use copulas to model the dependence among cyber incidents of different kinds and across various industries. \citet{xu_modeling_2018} use PRC's dataset to estimate time series models for the inter-arrival time and size of breaches and to model the dependence between them using copulas. \citet{kesan_when_2020} use a proprietary cyber incident dataset compiled by Advisen to study the key factors that drive the litigation following security incidents and the outcomes of lawsuits. \citet{kesan_empirical_2021} combine Advisen data with governmental information technology (IT) budget information to model the impact of IT spending on cyber risk. \citet{palsson_analysis_2020} use Advisen data and random forest classifiers and regressors to predict the legal and financial outcomes of cyber incidents based on characteristics of incidents and affected organizations. \citet{romanosky_examining_2016} builds linear models on Advisen data to study the factors influencing the financial impacts of cyber incidents on organizations. \citet{Eling2019} extract cyber-related incidents from an operational risk database, SAS OpRisk Global Data, and model the distribution of cyber losses in relation to incident and company characteristics.

It is worth noting that there are relatively few studies on entity-specific risk modeling because of the general lack of features associated with individual affected organizations in most of the available datasets, especially the ones in the public domain. For example, \citet{wheatley_addressing_2021} model the frequency and severity of data breaches at the industry level. In practice, according to \citet{romanosky_content_2019}, cyber insurance policies are often priced based on very few objective factors, including revenue and industry classification, in combination with expert judgment. From an underwriting perspective, this level of granularity may not be sufficient due to the significant variability of riskiness associated with organizations of similar sizes and in the same industry but with otherwise different risk characteristics. 

In addition, whereas dependence among different incident types is also often modeled at the industry level (see, for example, \citet{eling_copula_2018}), dependence at the entity level remains unknown. That is, does being susceptible to one type of incident suggest that the organization is prone to other types of incidents? This question relates to designing multiple-peril cyber insurance as discussed in \citet{chong_incident-specific_2025}. It is common practice for cyber insurance policies to cover multiple types of incidents, such as data breaches and ransomware attacks, and therefore a better understanding of entity-specific dependence is crucial.

\subsection{Cyber risk modeling with an enriched dataset}

In this study, we define a cyber incident as any event that compromises the confidentiality, integrity, or availability of entity-specific information systems, and we focus on modeling such incidents at the entity level to enrich existing datasets. Based on existing data and literature, we propose that cyber incident data, despite its scarcity in terms of number of records, can be enriched by entity-specific information of victim organizations, enabling a more thorough and detailed analysis of how the risk factors of an organization affect its cyber risk.

We make our best effort to collect cyber incident data from the public domain, which we then augment with additional data from a proprietary insurtech source that offers a rich set of features associated with individual organizations. This collective dataset enables us to build ML models for the assessment of entity-specific cyber risk and understand the occurrence and frequency of cyber incidents that an organization could possibly experience. The models are designed to answer the following questions:

\begin{enumerate}
    \item Compared with the conventional insurance risk factors used in practice, such as industry and revenue, do additional entity-specific organizational features offer an improved incident evaluation performance?
    \item Beyond capturing the occurrences of various incidents (i.e., whether an incident of a particular type would occur), can their frequencies (i.e., the number of incidents of a certain type) be evaluated reasonably well with the currently available data?
    \item At the entity level, when the firm-specific characteristics are given, is there any conditional dependence among the occurrences and frequencies of various kinds of incidents? 
\end{enumerate}

Our findings suggest that, besides revenue and industry, alternative entity-specific organizational insurtech-empowered features, such as the customer reviews of a business, indeed improve the performance of cyber risk models. The models also reveal that, with a rich set of entity-specific organizational features, the occurrence of various types of cyber incidents can be captured with reasonable accuracy, though evaluating their frequency remains a challenge. Finally, we find no evidence of entity-level dependence among the occurrences or frequencies of different types of cyber risks. We base this observation on the result that the classification and regression models with predictions on one type of incident being conditioned on another fail to prevail over the ones that treat individual incident types as independent. Note that this result does not challenge the intuition that firms with certain traits, such as strong cybersecurity, may experience lower occurrence and frequency across all types of cyber incidents. Rather, it suggests that the occurrences or frequencies of different types of incidents are likely to be conditionally independent, given the firm's characteristics.

The contributions of our work are as follows. The study offers a comprehensive review of cyber incident datasets in the public domain to facilitate future research in the field. That incident data is enriched by entity-specific organizational insurtech-empowered features, thus enabling risk and dependence modeling for individual organizations. The study further examines the effectiveness of entity-specific organizational features in capturing cyber incident occurrence and frequency and reveals a set of potential rating factors for cyber insurance. Lastly, the study suggests that dependence among different cyber incident types may not be present at the entity level.

The remainder of the paper is structured as follows. In Section~\ref{sec: data}, we introduce several publicly available cyber incident datasets, the insurtech data that offers additional entity-specific risk characteristics, and the comprehensive dataset that compiles data from those sources and that we use for the study. Section~\ref{sec:method} details the modeling methodology we adopt to answer the three aforementioned questions, and Section~\ref{sec:result} offers a summary of the performance of various models and the comparison among them. The feature importance results derived from the proposed models are described in Section~\ref{sec:FI}. In Section~\ref{sec:conclusion}, we discuss the results and implications of this study and conclude.

\section{Data}\label{sec: data}

To assess cyber risk at the individual-entity level, we leverage data and analytical tools made possible by recent advances in insurtech. Our approach incorporates entity-specific organizational digital footprints, social media activity, and other emerging data sources increasingly accessible through insurtech. This enables the development of more granular and dynamic risk assessments. We begin by identifying key outcome data related to past cyber incidents, such as incident type, followed by forward-looking risk factors that capture an entity's exposure to cyber threats, such as metrics associated with the entity's public image.

\subsection{Cyber incident data sources} \label{subsec:cid}

We offer a summary of the cyber incident datasets that exist in the public domain, including the aforementioned PRC data breach database and some others that are less commonly seen in academic studies. Despite variations in details captured by different data sources, all of the datasets we consider in this study contain at least some description of the nature of the incident, e.g., whether it is a data breach or a ransomware attack, the year the incident took place, and the entity affected by the incident.

PRC's database is a collection of publicly reported data breach incidents in the US (see \citet{PRC2025}). It gathers data security events disclosed by entities that are subject to the HIPAA Breach Notification Rule and/or state data breach notification laws. PRC pools the reported events from the US Department of Health and Human Services (HHS) and state law enforcement agencies, extracts additional information from the reported events, such as organization and breach type, and compiles the data from various sources into a standardized tabular format. The archive of incidents between 2005 and 2019, which contains more than 9,000 incidents, is freely available.\footnote{\url{https://privacyrights.org/sites/default/files/2020-01/PRC\%20Data\%20Breach\%20Chronology\%20-\%201.13.20.csv}.} Records of more recent incidents have to be purchased. Because PRC collects information from public sources, it is possible to manually and directly retrieve the recently disclosed incidents from HHS and states that publish such data, such as California and Oregon, but collecting data from those sources individually requires significant effort. 

The VERIS Community Database (VCDB) is a community-maintained public database of security incidents that follows the Vocabulary for Event Recording and Incident Sharing (VERIS) data model (see \citet{vcdb_repository}). The VERIS schema specifies the information to be recorded for each cybersecurity incident in standard formats, including victim demographics, incident description, responses, and outcomes. Actively maintained at the time of this writing, the database contains more than 10,000 records contributed by community members. However, because the records are voluntarily reported or collected, some potentially come from unverified sources and may have accuracy issues. The VERIS framework includes a confidence rating for the accuracy of each record. About 3,000 of the 10,000 records have a confidence rating of ``medium'' or ``high.''

The Center for International and Security Studies at Maryland (CISSM) maintains a Cyber Events Database (the CISSM database) (see \citet{Harry2018}). A research team from the University of Maryland's School of Public Policy scrapes websites that publish information on cybersecurity incidents, manually reviews the gathered data, and documents the incidents in a tabular format. The data collection is performed monthly, and at the time of this writing, the number of recorded incidents in this database is over 14,000.

The aforementioned three datasets are among the largest cyber incident datasets in the public domain. In addition, there are some smaller but notable ones. The Cybersecurity in Application, Research and Education Lab at Temple University maintains a Critical Infrastructure Ransomware Attacks dataset that collects data on publicly disclosed or reported ransomware attacks. It currently has more than 2,000 records (see \citet{Rege2022}). Researchers at the University of Queensland created a dataset of data breaches and ransomware attacks spanning a period of 15 years, starting in 2004. The dataset has over 1,000 recorded incidents (see \citet{Ko2020}). Have I Been Pwned, a dataset that keeps track of websites that have been breached and the amount and type of leaked data, currently includes data from more than 800 incidents (See \citet{Hunt2025}).

In addition to the structured datasets discussed above, another valuable source of information on historical security incidents is \citet{Doe2025}, which regularly publishes brief articles covering recent cyber incidents and other cybersecurity-related news, such as regulatory updates. Since 2008, the platform has published more than 35,000 articles, making it a rich repository of unstructured text data. Due to its unstructured nature, however, extracting meaningful insights requires advanced natural language processing (NLP) techniques. Inspired by our prior work in cyber literature collection and NLP, we apply a text analysis pipeline to extract structured information from these articles (see \citet{Zhang03042025}). Specifically, we utilize DeepSeek-R1 7B, a pretrained large language model (see \citet{deepseekai2025deepseekr1incentivizingreasoningcapability}), to identify and extract several key attributes from each article, including (1) whether the article describes a cyber incident, (2) the name of the affected entity, (3) the date of the incident if it is a recent incident, (4) whether the article is a follow-up report on an incident in the past, and (5) the type of the incident. This approach produced more than 11,000 identified incidents with known incident dates, incident types, and affected organizations. 

In this study, for each cyber incident dataset, we keep only the names of affected organizations, incident dates, and incident types. The originally documented incident types vary from dataset to dataset. For example, the CISSM database categorizes incidents into four groups: disruptive, exploitive, mixed, and undetermined. Disruptive and exploitive incidents are further categorized into 10 subgroups. In comparison, each incident in the VCDB is associated with one or more of six actions that led to the event, such as hacking and error-related causes, and one or more affected assets from a comprehensive list, such as database servers and network infrastructure. To unify these different approaches to incident classification, we follow the four categories proposed in \citet{kesan_empirical_2021}---i.e., data breach, privacy violation, extortion/fraud, and IT error---and map the existing set of categories used by each cyber incident dataset to them. Incidents that cannot be classified are labeled as \textit{other}.

Note that these data sources have overlapping records. For example, the mass data breach experienced by Equifax in 2017 appears in multiple aforementioned datasets. To create a single combined dataset for this study, we check for duplications under the assumption that an organization does not experience more than one incident of the same type on the same day, and duplicated records are removed. 

\subsection{Insurtech data} \label{sec: insurTechdata}

Traditional cyber insurance underwriting largely relies on conventional risk factors such as industry classification, business size, annual revenue, and responses to IT security surveys, as summarized by \citet{tsohou_cyber_2023, nurse_data_2020, romanosky_content_2019}. While such inputs offer a general overview, they often fall short of capturing the true nature and severity of an organization's cyber risk. In many cases, the information is either too generic or misaligned with the way attackers actually assess targets. For instance, survey-based questions such as ``How much does your company spend annually on IT security?'' may not yield accurate or meaningful answers. IT budgets are often approved at a broad entity-specific organizational level, making it difficult for respondents to isolate security-specific expenditures. Moreover, some policyholders may give incomplete or intentionally misleading responses, whether due to uncertainty, oversight, or misaligned incentives. As a result, survey-derived data may lack the precision and reliability needed for effective underwriting. As \citet{romanosky_content_2019} point out, the connection between potential cyber losses and the collected security information is unlikely to be quantitatively determined. Also, as \citet{tsohou_cyber_2023} suggest, the lack of an objective way to assess cyber risk contributes to the difficulties in assessing such risk, and the cyber insurance industry is in need of a risk assessment standard beyond the customized questionnaires that are commonly seen in current practice.  

To address these shortcomings, it is critical to incorporate more objective and externally observable indicators of cyber risk. Insurtech platforms now enable access to digital signals such as an organization's online footprint, social media activity, exposed credentials, domain registration history, unpatched software vulnerabilities, and public-facing infrastructure configurations. These external data points provide a more accurate and real-time reflection of an entity's risk posture from the perspective of potential attackers. 

For this study, we acquire a dataset of over 500 entity-specific organizational features from an insurtech platform, Carpe Data.\footnote{https://carpe.io/.} These features encompass a comprehensive range of business indicators, including geographic information, enterprise structure, operating status, customer review metrics, publicly disclosed business risk characteristics, classification segment tags, firmographics (e.g., size, revenue, employees), licensing categories, open hours statistics, and minority shareholding information. A detailed description of the insurtech dataset can be found in \citet{Quan15102024}. To maintain conciseness, we omit a comprehensive overview here and instead introduce relevant data components as they pertain to our discussion. {Briefly, we provide a category-level summary of the 527 insurtech features we use in this study in Appendix~\ref{app:notationD}, Table~\ref{tab:feature-summary-527}.}

Notably, this metadata is not derived from internal cyber incident records but from publicly accessible digital and operational attributes of the enterprise. Many of these features have not been examined in the cyber risk literature, and we shall later show that some of them---such as several key metrics associated with a firm's customer reviews---significantly affect the cyber risk of a firm.

\subsection{Assembled data}

The final datasets used for both multilabel classification and multi-output regression are constructed through a comprehensive data integration and processing pipeline, incorporating two primary sources: cyber incident data and insurtech data.

The cyber incident dataset is first compiled using the data sources described in Section~\ref{subsec:cid} and additional incident records from a proprietary source. It includes three features: \textit{COMPANY\_ID} (the unique company identifier), \textit{CASE\_TYPE\_LG} (the type of cyber incident), and \textit{ACCIDENT\_YEAR} (the year the incident occurred, ranging from 1903 to 2018). Following the removal of records with missing identifiers or years, and after applying a multistage deduplication process, the dataset is reduced to $66,215$ valid entries.

To prepare this data for modeling, the categorical incident type variable \textit{CASE\_TYPE\_LG} is converted into a multilabel format using one-hot encoding, resulting in five binary indicators corresponding to distinct types of cyber incidents: (1) privacy violation, (2) data breach, (3) extortion/fraud, (4) IT error, and (5) other. Each row in the cyber incident dataset corresponds to a single reported incident. To study how insurtech-empowered features relate to the occurrence and frequency of cyber incidents, we restructure the data at the firm-year level, where each observation represents a single firm in a single observed calendar year and consolidates all incidents experienced by that firm during that calendar year. Specifically, to obtain this data structure, the five binary indicators are aggregated accordingly at the firm-year level, collapsing multiple incident-level entries into a single firm-year summary. For instance, if a firm experienced two privacy violations and one IT error in 2019, its aggregated frequency vector becomes 
\[
\textit{Privacy Violation} = 2,\;
\textit{Data Breach} = 0,\;
\textit{Extortion/Fraud} = 0,\;
\textit{IT Error} = 1,\;
\textit{Other} = 0.
\]
This transformation ensures that the output variables reflect the frequency with which each incident type occurred within the year. After applying this transformation across all firms and calendar years, the dataset contains a total of 53,243 unique firm-year observations, each representing the full distribution of cyber incident types experienced by a company in a given calendar year. The aggregated frequency counts also allow us to construct the binary occurrence indicators by mapping any positive count to 1.

In parallel, the insurtech partner contributes a rich dataset of $528$ potential risk factors covering $31{,}782$ companies. To integrate the cyber incident records with the insurtech dataset, we align the two datasets at the firm-year level by merging observations through their shared company identifier: \textit{COMPANY\_ID}. However, due to limitations in historical data coverage, specifically the lack of company identifiers for incidents before the 2000s, the information for incidents occurring in the early 20th century is excluded during alignment. After retaining only firm-year pairs that appear in both datasets and removing identifier-related variables, the final modeling dataset contains 39,636 firm-year observations with 533 columns. Among these, 528 are features: 527 from the insurtech dataset and one from the cyber incident dataset (\textit{ACCIDENT\_YEAR}). The remaining five columns constitute the multilabel output variables.

Formally, for the multilabel classification problem, we define the dataset as
\[
\mathcal{D}_{\text{cls}} = \left\{(\mathbf{x}_i, \mathbf{y}_i)\right\}_{i=1}^{m} ,
\]
where each input \( \mathbf{x}_i \in \mathbb{R}^d \) represents a \( d \)-dimensional feature vector, and \( m \) denotes the total number of observations in the dataset. For insurtech-empowered features, the vector can be written as
\[
\mathbf{x}_i = (x_{i1}, x_{i2}, \dots, x_{id}), \quad i \in \{1, 2, \dots, m\} ,
\]
where $d=528$ and $m=39{,}636$. 

For the output variable space \( \mathcal{Y} = \{0,1\}^q \), each observation has a \( q \)-dimensional output variable vector \( \mathbf{y}_i \) from the label set, \( \mathcal{L} = \{L_1, L_2, \dots, L_q\} \), with each \( L_j \) representing a unique label, the index \( j \in \{1, 2, \dots, q\} \) preserving the order of labels across all observations. In our case, the label set \( \mathcal{L} = \{\text{Privacy Violation}, \text{Data Breach}, \text{Extortion/Fraud}, \text{IT Error}, \text{Other}\} \), where \( q =5\). The output variable vector \( \mathbf{y}_i \) serves as a binary indicator of the presence (1) or absence (0) of each \( L_j \)  for  \( i \)-th observation and can be written as
\[
\mathbf{y}_i = (y_{i1}, y_{i2}, \dots, y_{iq}), \quad i \in \{1, 2, \dots, m\}, 
\]
where $q=5$ and $m=39{,}636$.

Specifically, the \( j \)-th element of \( \mathbf{y}_i \) can be written as
\[
y_{ij} = 
\begin{cases}
1, & \text{if the label } L_j \text{ presents in \( i \)-th observation } \\
0, & \text{otherwise}
\end{cases} ,  \quad j\in \{1, 2, \dots, q\}.
\]
Hence, we can have multiple incident types present in a single firm-year observation.

Similarly, for the multi-output regression problem, we define the dataset as
\[
\mathcal{D}_{\text{reg}} = \{(\mathbf{x}_i, \mathbf{z}_i)\}_{i=1}^m ,
\]
where each output variable vector \( \mathbf{z}_i \in \mathbb{R}^q\) contains the count of cyber incidents in each of the \( q \) categories for firm \( i \) during a specific year, where the observation size $m$ remains consistent with the classification setting. The output dimensions are denoted by \( \mathcal{O} = \{O_1, O_2, \dots, O_q\} \), corresponding to the same categories as in the classification task. Thus, the output variable vector \( \mathbf{z}_i \) represents the number of times cyber incident type \( O_j \) occurred for the \( i \)-th observation. Specifically, the \( j \)-th element of \( \mathbf{z}_i \) can be written as
\[
z_{ij} \in \mathbb{R}, 
\quad i \in \{1, 2, \dots, m\}, \quad j \in \{1, 2, \dots, q\}.
\]

For both modeling tasks, the dataset is randomly partitioned into training and test subsets using a fixed split ratio of $80\%$ and $20\%$, respectively, ensuring consistent evaluation across methods.  

\section{Methodologies} \label{sec:method}

\subsection{Multilabel classification}\label{subsec:cls}

Multilabel classification is a supervised learning framework in which each observation can be simultaneously associated with multiple labels, rather than being limited to a single mutually exclusive category. This approach is particularly relevant in actuarial applications where complex relationships exist; for example, a single policyholder may hold multiple types of cyber insurance coverage with a company. In such cases, the traditional assumption of one label per observation does not hold. The multilabel setting introduces additional modeling challenges, including the need to account for dependencies among labels and to address issues related to label sparsity and class imbalance, both of which are common in real-world insurance datasets.

\subsubsection{Binary relevance}

Binary relevance (BR) (see \citet{tsoumakas2008multi}) decomposes the multilabel classification problem into $q$ independent binary classification tasks, one per label $L_j \in \mathcal{L} = \{L_1, L_2, \dots, L_q\}$. Given an input feature vector $\mathbf{x}_i \in \mathbb{R}^d$, BR assigns a binary output $\hat{y}_{ij} \in \{0,1\}$ to each label using a dedicated classifier $f_j : \mathbb{R}^d \to \mathbb{R}$. The prediction is thresholded by a label-specific parameter $\tau_j^* \in [0,1]$ as follows:
\[
\hat{y}_{ij} = \mathbb{I}\left(f_j(\mathbf{x}_i) \geq \tau_j^*\right), \quad j \in \{1, \dots, q\}.
\]
The complete prediction for observation $i$ is given by $\hat{\mathbf{y}}_i = (\hat{y}_{i1}, \dots, \hat{y}_{iq}) \in \{0,1\}^q$. While BR is simple and scalable, it assumes independence among labels, potentially limiting performance when label correlations exist.

\begin{algorithm}[!htbp]
\footnotesize
\caption{Binary Relevance with Joint Search for Multilabel Classification}
\label{alg:br_joint_search}
\SetAlgoLined
\DontPrintSemicolon

\KwIn{
    Feature matrix $\mathbf{X}_{\text{train}} \in \mathbb{R}^{r \times d}$; Label matrix $\mathbf{Y}_{\text{train}} \in \{0,1\}^{r \times q}$; Test feature matrix $\mathbf{X}_{\text{test}} \in \mathbb{R}^{s \times d}$; Test label matrix $\mathbf{Y}_{\text{test}} \in \{0,1\}^{s \times q}$; Label set $\mathcal{L} = \{L_1, \dots, L_q\}$; Base learner family $\mathcal{B}$ (e.g., $\text{LightGBM, Random Forest}$); Hyperparameter space $\Theta$ for the base learner; Threshold space $\mathcal{T}$; The number of cross-validation $K$; Loss function; Evaluation metric $F$
}

\KwOut{
    Label-specific thresholds $\boldsymbol{\tau}^* = \left(\tau_1^*, \dots, \tau_q^*\right)$; Predicted label matrix $\hat{\mathbf{Y}}_{\text{test}} \in \{0,1\}^{s \times q}$; Optimal hyperparameter $\theta^* \in \Theta$
}

\ForEach{$(\theta, \boldsymbol{\tau}) \in \Theta \times \mathcal{T}$}{

    Perform $K$-fold cross-validation on $\left(\mathbf{X}_{\text{train}}, \mathbf{Y}_{\text{train}}\right)$ \\

    \ForEach{validation fold}{
            Construct dataset: $\mathcal{D}_j \gets \left\{(\mathbf{x}_i, y_{ij})\right\}_{i \in \text{training folds}}$ for each label $j$\\
            Train base learner $f_{j,\theta} \in \mathcal{B}$ on $\mathcal{D}_j$ \\
            Predict validation scores: $\mathbf{p}_j^{\text{val}} \gets f_{j,\theta}\left(\mathbf{X}_{\text{val}}\right)$ \\
            Binarize predictions: $\hat{\mathbf{y}}_j^{\text{val}} \gets \mathbb{I}\left(\mathbf{p}_j^{\text{val}} \ge \tau_j\right)$\\
            Retrieve validation labels: $\mathbf{Y}_{\text{val},j}$ \\
        }

        Assemble: $\hat{\mathbf{Y}}_{\text{val}} \gets \left(\hat{\mathbf{y}}_1^{\text{val}}, \dots, \hat{\mathbf{y}}_q^{\text{val}}\right)$ \\
        Compute cross-validation loss: $\text{CVLoss}_{\text{fold}} \gets F\left(\hat{\mathbf{Y}}_{\text{val}}, \mathbf{Y}_{\text{val}}\right)$
}

Select best configuration: $\left(\theta^*, \boldsymbol{\tau}^*\right) \gets \arg\min_{(\theta, \boldsymbol{\tau})} \frac{1}{K} \sum \text{CVLoss}_{\text{fold}}(\theta, \boldsymbol{\tau})$

\ForEach{$L_j \in \mathcal{L}$}{
    Train $f_{j,\theta^*} \gets \mathcal{B}_{\theta^*}$ on $\left\{(\mathbf{x}_i, y_{ij})\right\}_{i=1}^{r}$, where $\mathbf{x}_i \in \mathbf{X}_{\text{train}},\ y_{ij} \in \mathbf{Y}_{\text{train}}$ \\
    Set $\tau_j^* \gets \tau_{j}$

}

\For{$i = 1$ \KwTo $s$}{
    Retrieve: $\mathbf{x}_i^* \in \mathbf{X}_{\text{test}}$ \\
    \For{$j = 1$ \KwTo $q$}{
        $\hat{y}_{ij}^* \gets \mathbb{I}\left(f_{j,\theta^*}(\mathbf{x}_i^*) \ge \tau_j^*\right)$
    }
    Set predicted label vector: $\hat{\mathbf{y}}_i^* \gets \left(\hat{y}_{i1}^*, \dots, \hat{y}_{iq}^*\right)$ \\
    Store row $i$: $\hat{\mathbf{Y}}_{\text{test}}[i] \gets \hat{\mathbf{y}}_i^*$
}

\Return{$\left(\boldsymbol{\tau}^*,\, \hat{\mathbf{Y}}_{\text{\upshape test}},\, \theta^*\right)$}
\end{algorithm}

Algorithm~\ref{alg:br_joint_search} outlines our modified BR approach for addressing the multilabel classification problem. In this framework, each label is treated as an independent binary classification task, allowing the use of any standard classification algorithm as the base learner. In our implementation, we experiment with multiple base classifiers for the function $\mathcal{B}$, including LightGBM (see \citet{ke2017lightgbm}) and random forest (see \citet{breiman2001random}). Both classifiers are tree-based models, providing the level of explainability essential for insurance ratemaking. LightGBM uses advanced, scalable algorithms that offer high flexibility, making it particularly well suited for industrial-scale applications. In contrast, while random forest typically demands more computational resources, it consistently delivers strong performance among traditional base learners, especially when applied to large datasets. The flexibility in selecting base classifiers not only allows us to evaluate the accuracy and efficiency of different learning algorithms but also facilitates testing of the dependency structure, while reducing variation attributable to the choice of base model. In addition, validation metric functions $F$ are formally defined and discussed in Appendix~\ref{app:notationB}. These measures provide a comprehensive view of model performance by capturing various aspects of classification quality across all labels.

\subsubsection{Classifier chains}

Classifier chains (CCs) (see \citet{read2011classifier}) extend BR by explicitly modeling label dependencies. Let $\boldsymbol{\sigma} = (\sigma_1, \dots, \sigma_q)$ denote a permutation of $\{1, \dots, q\}$ that specifies a fixed label ordering, where $\sigma_j$ represents the index of the label at position $j$ in the sequence. Since each order defines a distinct sequence of classifiers in the chain, this yields a total of $q!$ permutations, collectively denoted by $\Sigma$.

For the first label to predict, $\sigma_1$, the classifier $f_{\sigma_1}$ takes the feature vector $\mathbf{x}_i$ as input. For each subsequent classifier $f_{\sigma_j}$, the input consists of the original feature vector $\mathbf{x}_i$ and the predictions for the labels $\sigma_1$ through $\sigma_{j-1}$. The prediction for the dimension $j$ in the chain is defined as
\[
\begin{aligned}
\hat{y}_{i,\sigma_j} &= \mathbb{I}\left( f_{\sigma_j}(\mathbf{x}_i) \ge \tau_{\sigma_j}^* \right), \quad j = 1;\\
\hat{y}_{i,\sigma_j} &= \mathbb{I}\left( f_{\sigma_j}(\mathbf{x}_i, \hat{y}_{i,\sigma_1}, \dots, \hat{y}_{i,\sigma_{j-1}}) \ge \tau_{\sigma_j}^* \right), \quad j \in \{2, \dots, q\}.
\end{aligned}
\]

Algorithm~\ref{alg:cc_joint_search} outlines our modified CC approach for multilabel classification. In this framework, labels are predicted sequentially using a chain of classifiers, where each classifier incorporates previous labels' predictions as additional features, effectively modeling label dependencies. This chaining mechanism enables the model to capture label interdependence. However, its performance is often sensitive to the choice of the label order $\boldsymbol{\sigma}$.

\begin{algorithm}[!htbp]
\footnotesize
\caption{ Classifier Chain with Joint Search for Multilabel Classification}
\label{alg:cc_joint_search}
\SetAlgoLined
\DontPrintSemicolon

\KwIn{
Feature matrix $\mathbf{X}_{\text{train}} \in \mathbb{R}^{r \times d}$; Label matrix $\mathbf{Y}_{\text{train}} \in \{0,1\}^{r \times q}$; Test feature matrix $\mathbf{X}_{\text{test}} \in \mathbb{R}^{s \times d}$; Test label matrix $\mathbf{Y}_{\text{test}} \in \{0,1\}^{s \times q}$; Label set $\mathcal{L} = \{L_1, \dots, L_q\}$; Label order space $\Sigma$; Base learner family $\mathcal{B}$ (e.g., $\text{LightGBM, random forest}$); Hyperparameter space $\Theta$ for the base learner; Threshold space $\mathcal{T}$; The number of cross-validation $K$; Loss function; Evaluation metric $F$
}

\KwOut{
Optimal label order $\boldsymbol{\sigma}^*$; Label-specific thresholds $\boldsymbol{\tau}^* = \left(\tau_1^*, \dots, \tau_q^*\right)$; Predicted label matrix $\hat{\mathbf{Y}}_{\text{test}} \in \{0,1\}^{s \times q}$; Optimal hyperparameter $\theta^* \in \Theta$
}

\ForEach{$(\boldsymbol{\sigma}, \theta, \boldsymbol{\tau}) \in \Sigma \times \Theta \times \mathcal{T}$}{
    Perform $K$-fold cross-validation on $\left(\mathbf{X}_{\text{train}}, \mathbf{Y}_{\text{train}}\right)$ \\
    
    \ForEach{validation fold}{
        \For{$j = 1$ \KwTo $q$}{
            Construct dataset: $\mathcal{D}_{\sigma_j}^{\text{fold}} \gets \left\{(\tilde{\mathbf{x}}_i^{(j)}, y_{i,\sigma_j})\right\}_{i \in \text{training folds}}$, where $\tilde{\mathbf{x}}_i^{(j)} = \left(\mathbf{x}_i, \hat{y}_{i,\sigma_1}, \dots, \hat{y}_{i,\sigma_{j-1}}\right)$ are the augmented features; for $j=1$, we set $\tilde{\mathbf{x}}_i^{(1)} = \mathbf{x}_i$. \\
            Train base learner $f_{\sigma_j, \theta} \in \mathcal{B}$ on $\mathcal{D}_{\sigma_j}^{\text{fold}}$ \\
            Predict validation scores: $\mathbf{p}_{\sigma_j}^{\text{val}} \gets f_{\sigma_j, \theta}\left(\tilde{\mathbf{X}}_{\text{val}}^{(j)}\right)$ \\
            Binarize predictions: $\hat{\mathbf{y}}_{\sigma_j}^{\text{val}} = \mathbb{I}\left(\mathbf{p}_{\sigma_j}^{\text{val}} \ge \tau_{\sigma_j}\right)$ \\
            Retrieve validation labels: $\mathbf{Y}_{\text{val}, \sigma_j}$ \\
        }
        Assemble: $\hat{\mathbf{Y}}_{\boldsymbol{\sigma}}^{\text{val}} \gets \left(\hat{\mathbf{y}}_{\sigma_1}^{\text{val}}, \dots, \hat{\mathbf{y}}_{\sigma_q}^{\text{val}}\right)$ \\
        Compute fold loss: $\text{CVLoss}_{\boldsymbol{\sigma}, \theta, \boldsymbol{\tau}}^{\text{fold}} \gets F\left(\hat{\mathbf{Y}}_{\boldsymbol{\sigma}}^{\text{val}}, \mathbf{Y}_{\text{val}}\right)$
    }

    Compute average loss: $\text{CVLoss}_{\boldsymbol{\sigma}, \theta, \boldsymbol{\tau}} \gets \frac{1}{K} \sum \text{CVLoss}_{\boldsymbol{\sigma}, \theta, \boldsymbol{\tau}}^{\text{fold}}$
}

Select best configuration: $(\boldsymbol{\sigma}^*, \theta^*, \boldsymbol{\tau}^*) \gets \arg\min_{(\boldsymbol{\sigma},\theta,\boldsymbol{\tau})} \text{CVLoss}_{\boldsymbol{\sigma},\theta,\boldsymbol{\tau}}$

\For{$j = 1$ \KwTo $q$}{
    Train: $f_{\sigma_j^*, \theta^*} \gets \mathcal{B}_{\theta^*}$ on $\mathcal{D}_{\sigma_j^*}^{\text{train}}$\\
    Set $\tau_{\sigma_j^*} \gets \tau_{\sigma_j}$

}

\For{$i = 1$ \KwTo $s$}{
    Initialize: $\tilde{\mathbf{x}}_i^* \in \mathbf{X}_{\text{test}}$ \\
    \For{$j = 1$ \KwTo $q$}{
        Predict: $p_{\sigma_j^*}^* \gets f_{\sigma_j^*, \theta^*}\left(\tilde{\mathbf{x}}_i^*\right)$ \\
        Apply threshold: $\hat{y}_{i,\sigma_j^*}^* \gets \mathbb{I}\left(p_{\sigma_j^*}^* \ge \tau_{\sigma_j^*}^*\right)$ \\
        Augment: $\tilde{\mathbf{x}}_i^* \gets \left(\tilde{\mathbf{x}}_i^*, \hat{y}_{i,\sigma_j^*}^*\right)$
    }
    Set predicted label vector: $\hat{\mathbf{y}}_i^* \gets \left(\hat{y}_{i,\sigma_1^*}^*, \dots, \hat{y}_{i,\sigma_q^*}^*\right)$ \\
    Store: $\hat{\mathbf{Y}}_{\text{test}}[i] \gets \hat{\mathbf{y}}_i^*$
}

\Return{$\left(\boldsymbol{\sigma}^*,\, \boldsymbol{\tau}^*,\, \hat{\mathbf{Y}}_{\text{\upshape test}},\, \theta^*\right)$}
\end{algorithm}

\subsubsection{Multilabel classification trees}

Multilabel classification trees (MCTs) jointly predict multiple binary labels using a collection of decision trees. Each input $\mathbf{x}_i \in \mathbb{R}^d$ is associated with a vector of binary label $\mathbf{y}_i \in \{0,1\}^q$, where $q$ is the number of labels.

Unlike BR, which trains one independent model per label, MCTs use a single integrated model to predict all labels simultaneously. Internally, each tree in the ensemble is trained to optimize the average impurity reduction across labels, allowing the model to capture interlabel dependencies. \citet{QuanValdez} provide a comprehensive discussion of multilabel (multivariate) tree-based models. In this study, we implement the MCT approach using the Python library \textit{RandomForestClassifier}, which ensembles multiple multilabel trees trained on random subsets of the data. Algorithm~\ref{alg:rf_multilabel_joint} outlines our modified MCT approach for multilabel classification.

\begin{algorithm}[!htbp]
\footnotesize
\caption{Multilabel Classification Trees with Joint Search}
\label{alg:rf_multilabel_joint}
\SetAlgoLined
\DontPrintSemicolon

\KwIn{
    Feature matrix $\mathbf{X}_{\text{train}} \in \mathbb{R}^{r \times d}$; Label matrix $\mathbf{Y}_{\text{train}} \in \{0,1\}^{r \times q}$; Test feature matrix $\mathbf{X}_{\text{test}} \in \mathbb{R}^{s \times d}$; Test label matrix $\mathbf{Y}_{\text{test}} \in \{0,1\}^{s \times q}$; Label set $\mathcal{L} = \{L_1, \dots, L_q\}$; Base learner $\mathcal{B}$ ($\text{RandomForestClassifier}$); Hyperparameter space $\Theta$ for the base learner; Threshold space $\mathcal{T}$; The number of cross-validation $K$; Loss function; Evaluation metric $F$
}
\KwOut{
    Optimal thresholds $\boldsymbol{\tau}^* = (\tau_1^*, \dots, \tau_q^*)$; Predicted label matrix $\hat{\mathbf{Y}}_{\text{test}} \in \{0,1\}^{s \times q}$; Optimal hyperparameter $\theta^* \in \Theta$
}

\ForEach{$(\theta, \boldsymbol{\tau}) \in \Theta \times \mathcal{T}$}{
    Perform $K$-fold cross-validation on $(\mathbf{X}_{\text{train}}, \mathbf{Y}_{\text{train}})$ \\
    \ForEach{validation fold}{
        Train base learner $f_{\theta} \in \mathcal{B}$ on training folds \\
        Predict validation scores: $\mathbf{P}^{\text{val}} \gets f_{\theta}(\mathbf{X}_{\text{val}})$ \\
        Binarize predictions:
        $\hat{\mathbf{Y}}^{\text{val}} \gets \mathbb{I}\left(\mathbf{P}^{\text{val}} \ge \boldsymbol{\tau}\right)$\\
        Retrieve validation labels: $\mathbf{Y}_{\text{val}}$ \\
         Compute cross-validation loss: $\text{CVLoss}_{\text{fold}} \gets F\left(\hat{\mathbf{Y}}_{\text{val}}, \mathbf{Y}_{\text{val}}\right)$
    }
}

 Select best configuration: $(\theta^*, \boldsymbol{\tau}^*) \gets \arg\min_{(\theta, \boldsymbol{\tau})} \frac{1}{K} \sum \text{CVLoss}_{\text{fold}}(\theta, \boldsymbol{\tau})$

Train final model: $f_{\theta^*} \gets \mathcal{B}_{\theta^*}$ on $(\mathbf{X}_{\text{train}}, \mathbf{Y}_{\text{train}})$ \\
Set $\boldsymbol{\tau}^* \gets (\tau_1, \dots, \tau_q)$

\For{$i = 1$ \KwTo $s$}{
    Retrieve: $\mathbf{x}_i^* \in \mathbf{X}_{\text{test}}$ \\
    Predict all labels jointly: $\mathbf{p}_i^* \gets f_{\theta^*}(\mathbf{x}_i^*)$ \\
    Apply vectorized thresholding: $\hat{\mathbf{y}}_i^* \gets \mathbb{I}\left(\mathbf{p}_i^* \ge \boldsymbol{\tau}^*\right)$ \\
    Store row $i$: $\hat{\mathbf{Y}}_{\text{test}}[i] \gets \hat{\mathbf{y}}_i^*$
}

\Return{$\left(\boldsymbol{\tau}^*,\, \hat{\mathbf{Y}}_{\text{\upshape test}},\, \theta^*\right)$}
\end{algorithm}

\subsubsection{Classification performance evaluation metrics}\label{subsec:cls_metrics}

In our study, multilabel classification introduces structural complexities, such as label co-occurrence patterns and imbalanced label distributions, that limit the effectiveness of traditional single-label evaluation metrics. As a result, specialized multilabel performance metrics have been developed to evaluate model performance appropriately. 

Among the commonly used evaluation metrics for multilabel classification are the micro-F1, macro-F1, weighted-F1, and sample-F1 scores, each emphasizing different aspects of model behavior. \citet{hinojosa2024performance} compare the F1 variants and demonstrate that metric selection can have a significant effect on evaluation outcomes, particularly under label imbalance. Specifically, the weighted-F1 score further compensates for label imbalance by weighting each label's contribution based on its prevalence in the dataset.

In addition to F1-based metrics, we also use the Jaccard index, which measures set-level similarity, and the Hamming loss, which quantifies the proportion of misclassified labels (see \citet{ganda2018survey}). Collectively, these metrics offer a comprehensive and robust framework for evaluating multilabel classification models under varying label distributions. Appendix~\ref{app:notationB} details the evaluation metrics we use in this study.

\subsection{Multi-output regression}\label{subsec:reg}

Multi-output regression is a supervised learning approach in which each observation is associated with multiple continuous outcomes. By jointly modeling these correlated outcomes, multi-output regression enables more coherent and efficient estimation, enhances predictive accuracy, and supports integrated decision-making, particularly valuable for insurers managing multifaceted risk exposures. Given the structural similarity between multi-output regression and multilabel classification, we omit a separate algorithmic description and instead focus on highlighting the key differences between the two frameworks in the discussion that follows.

\subsubsection{Multi-output regressor} 

The multi-output regressor (MOR) treats a multi-output regression problem as a set of $q$ independent single-output regression tasks, one for each output dimension $O_j \in \mathcal{O} = \{O_1, O_2, \dots, O_q\}$. Given an input feature vector $\mathbf{x}_i$, the MOR approach constructs a separate regressor $f_j: \mathbb{R}^d \rightarrow \mathbb{R}$ for each output dimension. The regressors are trained independently but in parallel, and the prediction for the dimension $j$ is defined as
\[
\hat{z}_{ij} = f_j(\mathbf{x}_i), \quad j \in \{1, \dots, q\}.
\]The complete prediction for observation $i$ is then expressed as $\hat{\mathbf{z}}_i = (\hat{z}_{i1}, \dots, \hat{z}_{iq}) \in \mathbb{R}^q$. This approach, commonly referred to as a problem transformation method, offers flexibility by enabling the use of any standard regression algorithm as the base learner. However, it assumes conditional independence among outputs given the input, which may hinder performance when output dependencies are informative. In our implementation, we test multiple base regressors $\mathcal{B}$, including LightGBM and random forest. We formally define the evaluation metrics used to assess model performance across all dimensions in Appendix~\ref{app:notationC}.

\subsubsection{Regressor chain}

The regressor chain (RC) (see \citet{spyromitros2012multi}) is inspired by chain-based strategies originally developed for multilabel classification (see \citet{read2011classifier}), which incorporate sequential dependency modeling among output variables. Similarly to CCs in the multilabel setting before, RC defines a permutation $\boldsymbol{\sigma} = (\sigma_1, \dots, \sigma_q)$ of the output indices $\{1, \dots, q\}$, determining the fixed ordering of output dimensions within the chain. Except for the regressor $f_{\sigma_1}$, which corresponds to the first label $\sigma_1$ and requires only the feature vector $\mathbf{x}_i$ as input, for each succeeding regressor \( f_{\sigma_j} \), the input is formed by augmenting the original feature vector \( \mathbf{x}_i \) with the predicted values of the preceding outputs in the chain, and it returns a prediction for the output dimension indexed by \( \sigma_j \):
\[
\begin{aligned}
\hat{z}_{i,\sigma_j} &= f_{\sigma_j}(\mathbf{x}_i), \quad j = 1; \\
\hat{z}_{i,\sigma_j} &= f_{\sigma_j}(\mathbf{x}_i, \hat{z}_{i,\sigma_1}, \dots, \hat{z}_{i,\sigma_{j-1}}), \quad j \in \{2, \dots, q\}.
\end{aligned}
\]
This chaining mechanism also allows the model to take advantage of the dependencies between the outputs. However, similar to its classification counterpart, RC is sensitive to the specified output ordering $\boldsymbol{\sigma}$. Furthermore, in scenarios where output dependencies are weak or negligible, the chaining mechanism may even introduce instability in model performance or additional computational burdens.

\subsubsection{Multi-output regression trees}

Multi-output regression trees (MRTs) extend traditional decision trees to simultaneously predict multiple continuous outputs.  Rather than training separate models for each output, MRTs construct a unified tree structure that partitions the input space to minimize a multivariate loss function. This formulation allows the model to account for correlations among outputs, and for each input \( \mathbf{x}_i \in \mathbb{R}^d \), the model estimates a vector \( \mathbf{z}_i \in \mathbb{R}^q \).

In this study, we implement the MRT approach with \textit{RandomForestRegressor} to construct a unified ensemble model. However, as the number of output dimensions increases ($q \gg 1$), evaluating multivariate splits becomes increasingly computationally intensive, which may hinder scalability.

\subsubsection{Regression performance evaluation metrics}\label{subsec:reg_metrics}

Compared with single-output regression tasks, where standard metrics such as mean squared error (MSE) or R-squared (R\textsuperscript{2}) are sufficient, multi-output regression requires evaluation metrics that account for the joint behavior of multiple continuous outputs. In this study, we evaluate model performance using the following six metrics: average MSE (aMSE), average RMSE (aRMSE), average relative RMSE (aRRMSE), average correlation coefficient (aCC), and global Euclidean distance (EU\_DIST), as described in Appendix~\ref{app:notationC}. These metrics collectively capture different dimensions of model performance, including error magnitude, scale-adjusted accuracy, and joint prediction consistency (see \citet{borchani2015survey}). 

\section{Data-driven analysis and findings} \label{sec:result}

In our empirical study, the classification task corresponds to capturing the occurrence of different types of cyber incidents, analogous to claim occurrence modeling in the insurance industry. Meanwhile, the regression task focuses on estimating the frequency or expected count of incidents, consistent with frequency modeling commonly used to assess actuarial pricing.

\subsection{Conventional data vs. insurtech-enriched data}

To evaluate the efficacy of entity-specific organizational features derived from insurtech on both cyber incident occurrence and frequency, we perform a comparative analysis using two datasets varying features with the same label: an existing conventional cyber incident dataset (D1) containing only the most commonly used rating factors, including industry classification and annual revenue, and an insurtech-enriched dataset (D2) that incorporates additional external entity-specific organizational features obtained from various platforms (with details provided in Section~\ref{sec: insurTechdata}). These enriched features provide an entity-specific representation of an organization's cyber risk posture. Note that the features in D1 are a subset of those in D2. 

The conventional dataset (D1) is evaluated using three representative models for each task, while the enriched dataset (D2) is evaluated using five model configurations to enable a broader assessment of both occurrence and frequency within the enriched data setting. Specifically, for classification, we implement BR using random forest and LightGBM classifiers extended via \textit{MultiOutputClassifier} (RF\_MOC and LGBM\_MOC), and implement CC using the same base classifiers (namely, RF\_CC and LGBM\_CC). We also incorporate MCTs, denoted as MLRF, which leverage tree ensembles to simultaneously predict multiple labels within a unified classification framework. For regression, we implement MOR using random forest and LightGBM regressors wrapped with \textit{MultiOutputRegressor} (namely, RF\_MOR and LGBM\_MOR), where each output is modeled independently without capturing interdependencies. To explore potential dependencies among outputs, we also implement RC with the same base regressors (namely, RF\_RC and LGBM\_RC). Additionally, we use MRTs, termed MORF, to jointly predict all output variables.

To facilitate a comprehensive comparison across datasets and model architectures, we construct comparative heatmaps to visualize the model performances across both classification and regression tasks. In our experiments, we evaluate model performance using two separate sets of evaluation metrics mentioned in Appendix~\ref{app:notationB} and Appendix~\ref{app:notationC}. To ensure consistency and facilitate comparison, evaluation metric values within each task are normalized across models to generate the heatmap's color gradient (darker blue tones reflect better model performance, in contrast to red tones indicating weaker results), while the original evaluation metric values are displayed within the heatmap cells to retain precise quantitative detail. Therefore, each row of the heatmap matrix represents a specific model, while each column corresponds to an evaluation metric, allowing for a systematic comparison of model performance across various architectures and dataset configurations.

\begin{figure}[!htbp]
    \centering
    \includegraphics[width=0.9\textwidth]{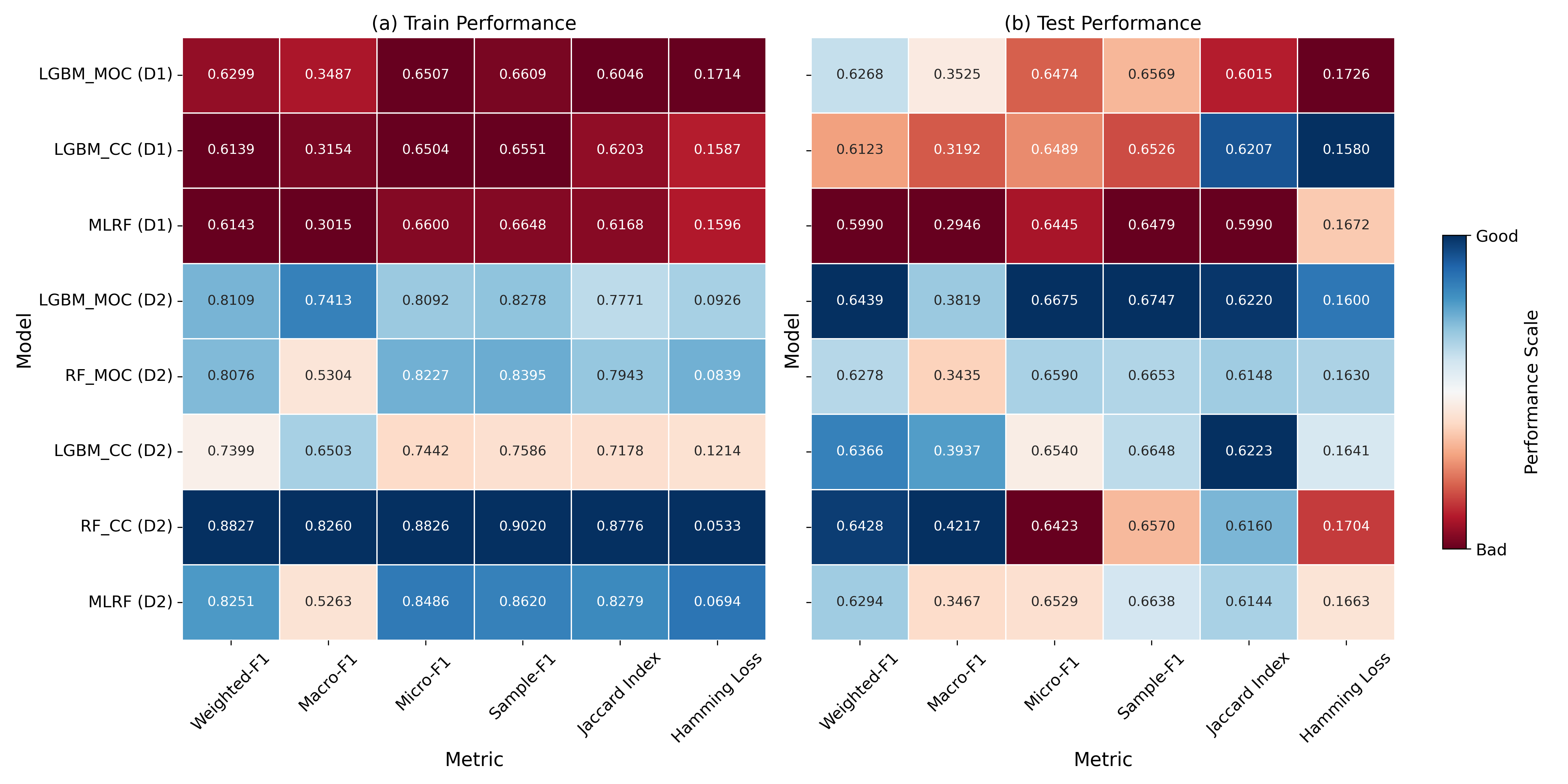}
    \caption{Multilabel classification model performance.}
    \label{fig:classification_model_performance}
\end{figure}

For cyber incident occurrence evaluation, we use six metrics: weighted-F1, macro-F1, micro-F1, and sample-F1 scores, as well as the Jaccard index and Hamming loss. In our analysis, we focus on weighted-F1 and Hamming loss as the most informative indicators of model performance (with details provided in Section~\ref{subsec:cls_metrics}). As shown in Figure~\ref{fig:classification_model_performance}, models trained on D2 consistently outperform those trained on D1 across both weighted-F1 and Jaccard index. Weighted-F1 scores on D2 reach values above 0.88 in the training set and over 0.64 in the test set, while the same metrics on D1 remain below 0.63. Additionally, the Jaccard index improves from an average of less than 0.61 on D1 to consistently exceeding 0.61 on D2. These results demonstrate that entity-specific organizational features from insurtech capture critical information that improves the model's ability to correctly identify multiple types of incidents, leading to more accurate and reliable performance.

\begin{figure}[!htbp]
    \centering
    \includegraphics[width=0.9\textwidth]{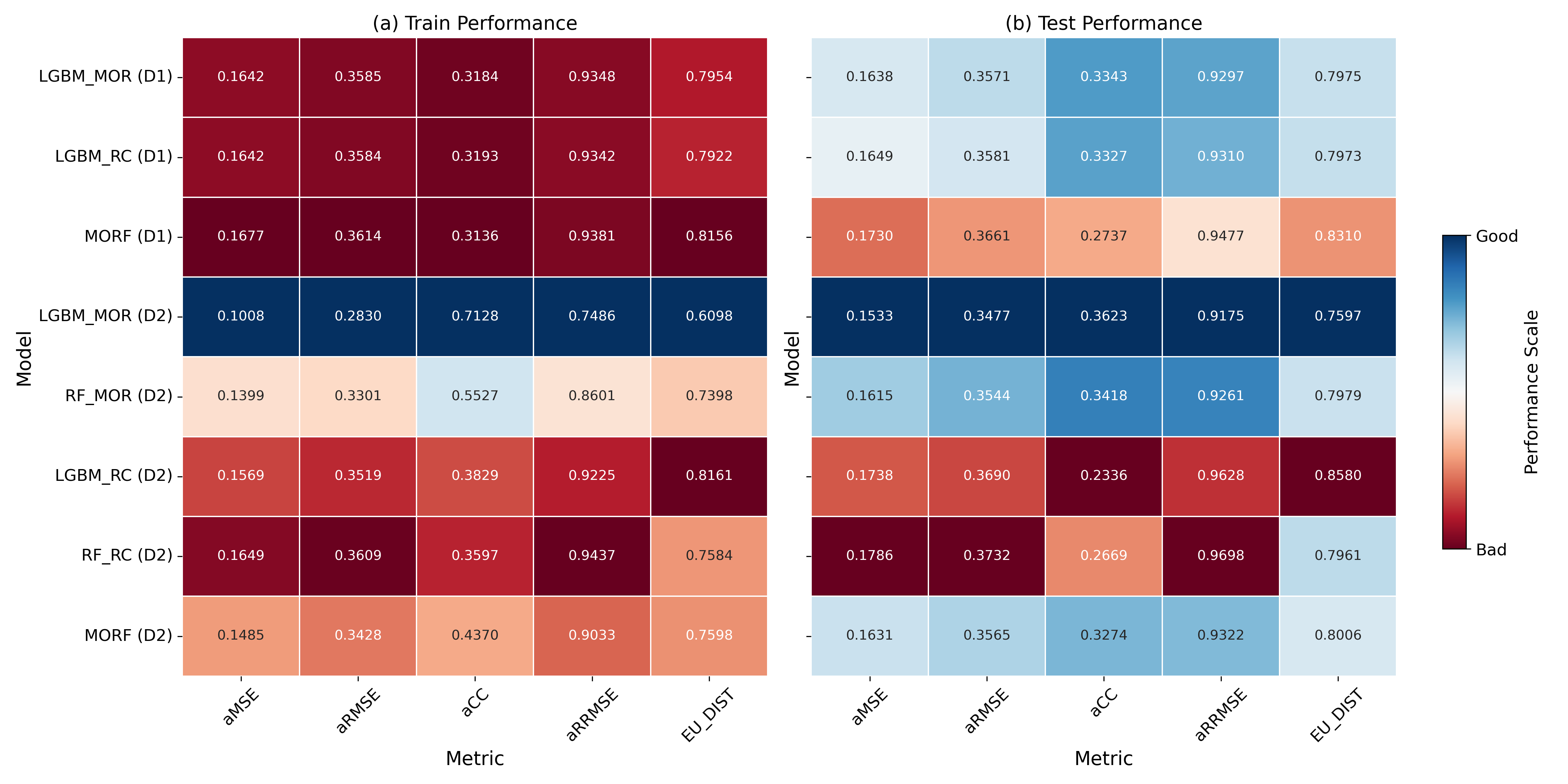}
    \caption{Multi-output regression model performance.}
    \label{fig:regressionn_model_performance}
\end{figure}

For cyber incident frequency evaluation, we assess model performance using aMSE, aRMSE, aCC, aRRMSE, and EU\_DIST, as summarized in Section~\ref{subsec:reg_metrics}. Overall, the enriched dataset (D2) provides substantial improvements in frequency estimation performance, particularly for models without dependency structures, as illustrated in Figure~\ref{fig:regressionn_model_performance}. Specifically, models trained on D2 consistently achieve lower aRMSE and aMSE values, with aRMSE decreasing from approximately 0.36 on D1 to as low as 0.28 on D2 in training, and the corresponding test set values follow a similar trend. Likewise, the aMSE values on D2 decrease below 0.11 in the train set and around 0.16 in the test set, compared with the higher values observed on D1. However, these improvements are not consistent across all model types, as certain dependency-based models, such as RC, exhibit degraded performance under D2. This highlights that misspecification in the dependency structure, especially when using high-dimensional entity-specific organizational features, can lead to overfitting on the training data. As a result, even with insurtech-enriched data, frequency estimates may be adversely affected. 

These findings collectively demonstrate that insurtech-enriched entity-specific organizational features play a pivotal role in improving occurrence modeling and frequency estimation by providing more informative and context-specific inputs.

\subsection{Metric-specific model performance}

On the insurtech-enriched dataset (D2), LGBM\_MOC achieves the highest weighted-F1 score (0.6439), demonstrating superior overall classification performance after adjusting for label imbalance. This indicates that the model performs effectively across the entire label distribution, including both frequent and infrequent classes. RF\_CC attains a very similar weighted-F1 score (0.6428); however, it exhibits the highest Hamming loss (0.1704) among the D2 models, reflecting a greater average number of label-wise prediction errors per instance. In contrast, LGBM\_MOC attains the second-lowest Hamming loss (0.1600), suggesting more conservative and precise label-wise predictions. This trade-off highlights the importance of evaluating multilabel models across complementary metrics: Weighted-F1 reflects performance relative to class distribution, whereas Hamming loss penalizes scattered misclassifications and overpredictions.

For frequency estimation, LGBM\_MOR achieves the best performance across all metrics. On D2, this model records an aRMSE of 0.2830 (train) and 0.3477 (test), along with an aMSE of 0.1008 (train) and 0.1533 (test), representing substantial reductions in average prediction errors compared to all D1 counterparts. The model also attains aCC values of 0.7128 (train) and 0.3623 (test), indicating improved alignment between predicted and observed incident frequencies. Similarly, the MORF model on D2 demonstrates lower aRMSE and aMSE values compared with its D1 version, confirming that entity-specific organizational insurtech-empowered features significantly enhance accuracy for models without explicit dependency structures. However, RC models show the opposite trend: Both RF\_RC and LGBM\_RC experience an increase in aRMSE (up to 0.3732 in the test set) and a decrease in aCC (as low as 0.2336) on D2, indicating both higher prediction errors and weaker alignment with observed frequencies. Although LGBM\_MOR shows consistently superior results across all metrics, the performance ranking of other models differs across different metrics. This lack of consistency illustrates that no single metric fully reflects model quality, highlighting the importance of multimetric evaluation, especially under insurtech-enriched feature settings.

\subsection{Model architecture}

Although classification model architectures differ in their capacity to capture interlabel dependencies, their overall impact on model performance is relatively limited within the context of our study. Architectures such as CC (e.g., RF\_CC, LGBM\_CC) and MLRF are explicitly constructed to capture interlabel dependencies and facilitate joint optimization over multiple output variables. Nonetheless, empirical results reveal that these structurally sophisticated models do not consistently yield superior model performance compared with simpler alternatives. Notably, LGBM\_MOC, a relatively straightforward model based on binary relevance with multi-output adaptation, attains the highest weighted-F1 score (0.6439) on D2, outperforming LGBM\_CC (0.6366) and MLRF (0.6294). 

In addition, for frequency estimation tasks, the dependency architecture has an inferior effect. Our results reveal that RC models exhibit instability under D2, where enriched entity-specific organizational insurtech features increase data complexity. RC models explicitly capture sequential output dependencies, incorporating the prediction order as a tunable hyperparameter. While, in theory, optimizing this order can improve model performance, the factorial growth of possible permutations with increasing label dimensionality makes exhaustive tuning computationally expensive. Empirically, both RF\_RC and LGBM\_RC demonstrate elevated aMSE and aRMSE, reaching up to 0.1786 and 0.3732 on the test set, alongside decreased CC values as low as 0.2336. These outcomes suggest that the interaction between insurtech-enriched features and dependency structures introduces additional system complexity, magnifying the sensitivity to output ordering and increasing the susceptibility to overfitting. Alternatively, dependency-free models such as LGBM\_MOR and MORF consistently achieve lower errors and greater prediction stability under D2, demonstrating the advantages of simpler architectures in effectively leveraging enriched entity-specific organizational features for reliable frequency estimation.

Therefore, in our cyber incident dataset, dependencies between different types of cyber incidents appear to be weak or nonexistent. Nevertheless, model architecture remains a critical consideration, as effectively leveraging enriched data requires selecting structures that balance model complexity with stability, especially in high-dimensional feature spaces.

\section{Feature importance} \label{sec:FI}

With over 500 features in the insurtech-enriched dataset, interpreting the factors driving model performance becomes challenging. To enhance the interpretability of the results, we identify the top 20 most influential features that contribute to the model performance of the multilabel classification and multi-output regression models described in Section~\ref{subsec:cls} and Section~\ref {subsec:reg}. To achieve this, we apply several widely used feature importance techniques, aiming to uncover key features and provide insights into potential cyber risk factors captured by the insurtech data.

\subsection {Feature importance techniques}

It is important to emphasize that feature importance techniques provide insight into how features influence a model's predictions, but they do not imply causal relationships. The observed associations reflect the model's internal mechanics rather than underlying cause-and-effect dynamics. Establishing causality would require a separate, rigorous analysis using formal causal inference methods, supported by domain expertise and possibly experimental or quasi-experimental designs.

Nonetheless, feature importance techniques serve as valuable tools for interpreting complex ML models, offering a window into the so-called ``black box.'' However, each technique has its own limitations, such as sensitivity to feature correlation, model bias, or instability under resampling, which can affect interpretability. To mitigate those concerns and improve the robustness of our findings, we apply multiple feature importance techniques: impurity-based importance, permutation importance, and SHapley Additive exPlanations (SHAP). This ensemble approach enables a more comprehensive and reliable identification of the key features driving the model's predictive performance.

\subsubsection {Impurity-based feature importance}

Impurity-based feature importance, often referred to as mean decrease impurity, introduced by~\citet{breiman2002dmi}, quantifies the influence of each input feature by aggregating its contributions to reductions in a chosen impurity criterion during the training process. For example, in random forest classification, the two predominant impurity criteria are the Gini index  (see \citet{lerman1984gini}) and information gain (see \citet{kent1983information}). The Gini index, widely adopted because of its computational efficiency, measures the probability of misclassification and reflects class heterogeneity. Alternatively, information gain, derived from entropy, serves as a more discriminative impurity measure by quantifying the expected reduction in uncertainty after a split. Compared with the Gini index, information gain often exhibits greater sensitivity to informative features in high-dimensional spaces, where many features may be irrelevant or redundant. By assigning minimal importance to uninformative features, it encourages sparsity in feature selection. However, its higher computational cost can limit scalability in large ensembles. Unlike classification trees that assess class purity, regression trees assess impurity in terms of prediction error. Specifically, reductions in MSE are used to evaluate split quality, and each decrease in output variance is attributed to the corresponding partitioning feature. In LightGBM, feature importance is frequently assessed using gain and split count. Gain measures the total reduction in the loss function resulting from splits involving a given feature, while split count indicates how often the feature is used for partitioning across all decision nodes. Unlike classical information gain, which is based on entropy reduction in classification tasks, LightGBM gain derives from improvements in the model-specific loss function (e.g., MSE or log loss), making it applicable to both regression and classification.

\subsubsection{Permutation feature importance}

Permutation feature importance, introduced by \citet{breiman2001random}, and formalized by \citet{fisher2019all}, is a method to evaluate the contribution of a feature by measuring the change in model performance after randomly permuting that feature's values. This approach maintains the original model and prediction process but introduces noise to one feature at a time to observe the effect on predictive accuracy. The difference in the evaluation metric before and after permutation indicates the importance of the feature. The method is model-agnostic, applicable to both classification and regression tasks, and supports a consistent assessment framework across different algorithms. It is particularly useful for interpreting complex models such as ensembles or neural networks. However, it may be sensitive to feature correlation, as highly correlated features can mask each other's contribution during permutation.

\subsubsection{SHapley Additive exPlanations feature importance}

SHAP, introduced by \citet{lundberg2017unified}, interprets model predictions by computing Shapley values derived from cooperative game theory. It explains each individual prediction by assigning every feature a local importance value based on its marginal contribution, averaged over all possible subsets of input features. This method ensures a fair and theoretically consistent allocation of feature contributions, satisfying key properties such as local accuracy, consistency, and missingness. To assess global feature importance, SHAP aggregates the absolute Shapley values across all observations, yielding a comprehensive ranking of features by their average contribution to model output. Compared with impurity-based and permutation-based approaches, SHAP captures interaction effects and the direction of influence (positive or negative), though it often requires greater computational resources. For tree-based models such as random forest and LightGBM, SHAP values can be computed efficiently using TreeExplainer (see  \citet{lundberg_local_2020}), a popular explainer tailored to ensemble tree models. TreeExplainer leverages the structure of decision trees to reduce the computational complexity from exponential to polynomial time, enabling exact and fast computation of Shapley values.

\subsection{Entity-level feature contributions to cyber incident occurrence and frequency}

To examine how entity-specific organizational features influence the occurrence of cyber incident types, we analyze feature importance across five model configurations: LGBM\_MOC, LGBM\_CC, RF\_MOC, RF\_CC, and MLRF. In addition, we extend this analysis to frequency estimation by evaluating feature importance across five regression model configurations: LGBM\_MOR, LGBM\_RC, RF\_MOR, RF\_RC, and MORF. For both tasks, we apply three different feature importance techniques---impurity-based importance, permutation importance, and SHAP---to provide a comprehensive evaluation of influential entity-specific organizational features.

\begin{figure}[!htbp]
    \centering
        \includegraphics[width=\textwidth]{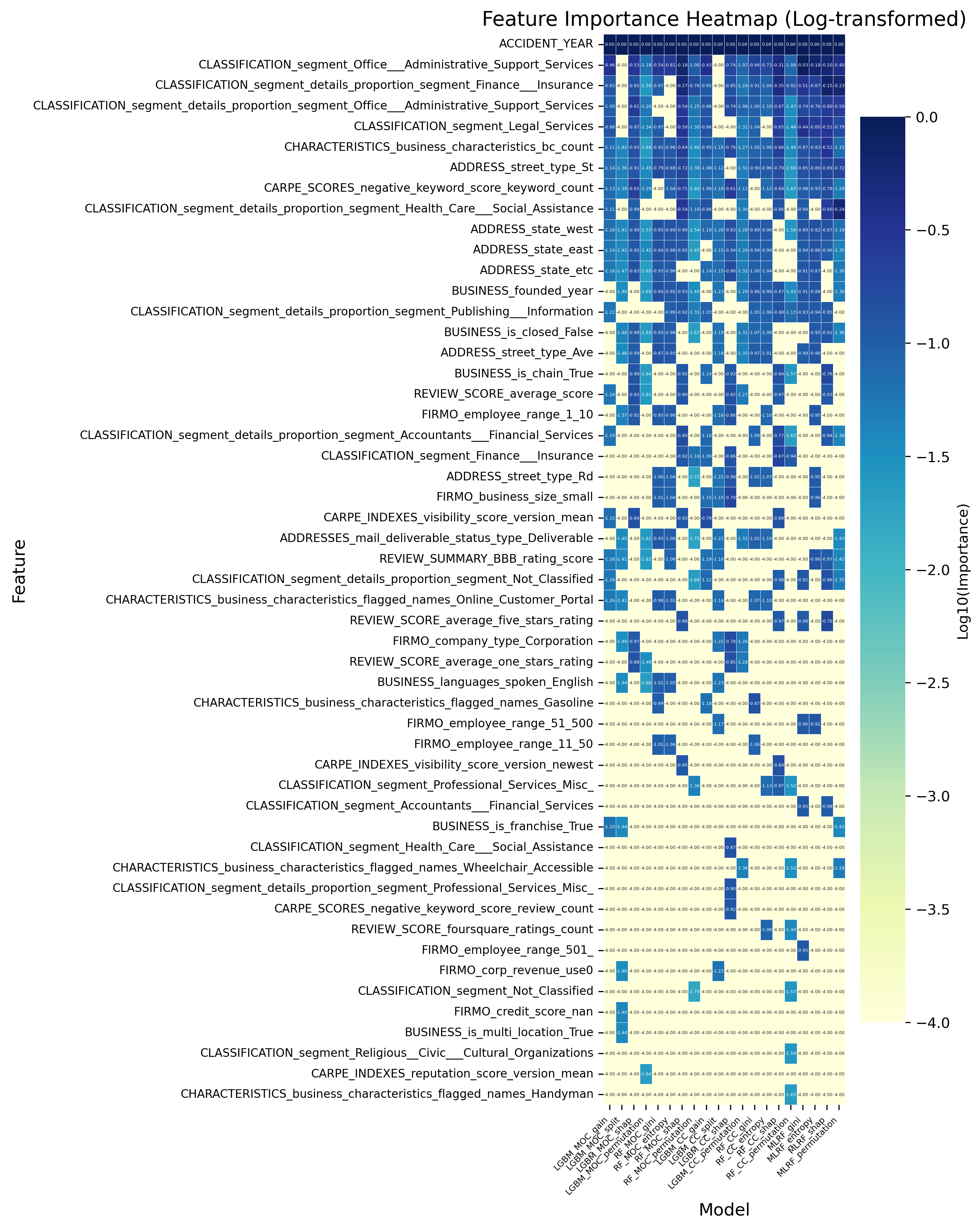}
        \caption{Log-transformed feature importance scores across various {classification} models.}
        \label{fig:importance_heatmap_cls}
\end{figure}

\begin{figure}[!htbp]
        \includegraphics[width=\textwidth]{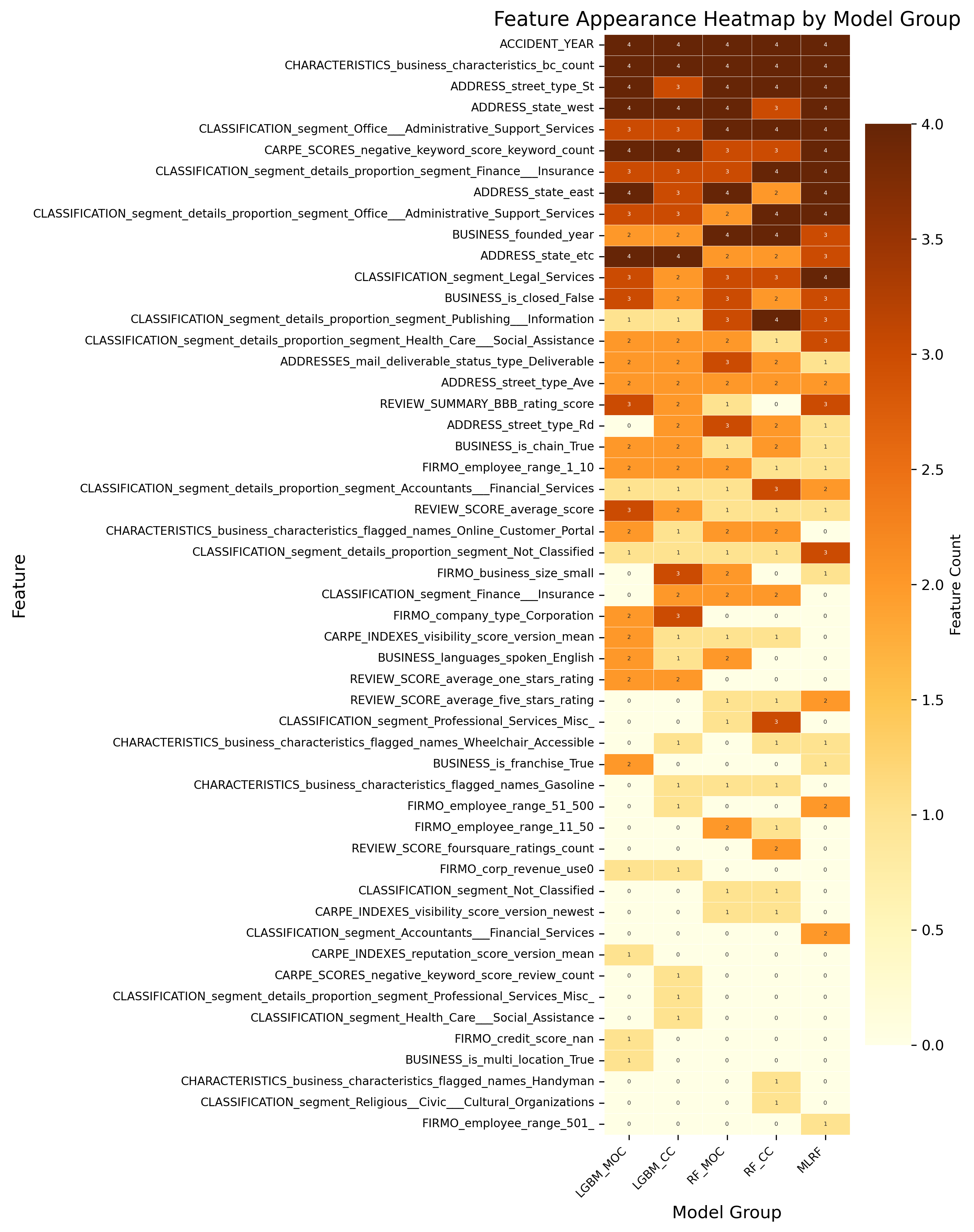}
        \caption{Count of top important feature appearance across various {classification} models.}
        \label{fig:feature_count_heatmap_cls}
\end{figure}

\begin{figure}[!htbp]
    \centering
        \includegraphics[width=\textwidth]{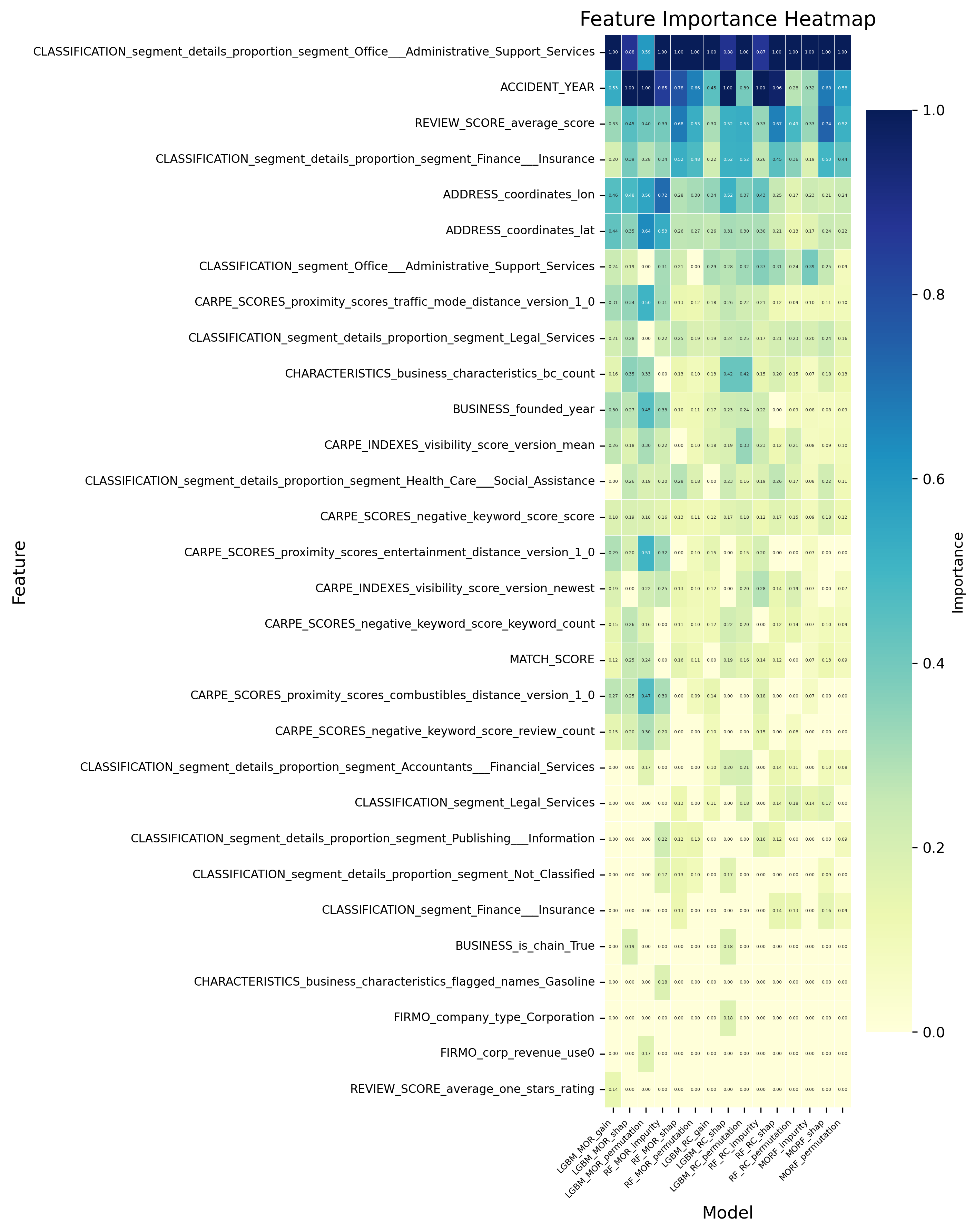}
        \caption{Feature importance scores across various {regression} models.}
        \label{fig:importance_heatmap_reg}
\end{figure}

\begin{figure}[!htbp]
        \includegraphics[width=\textwidth]{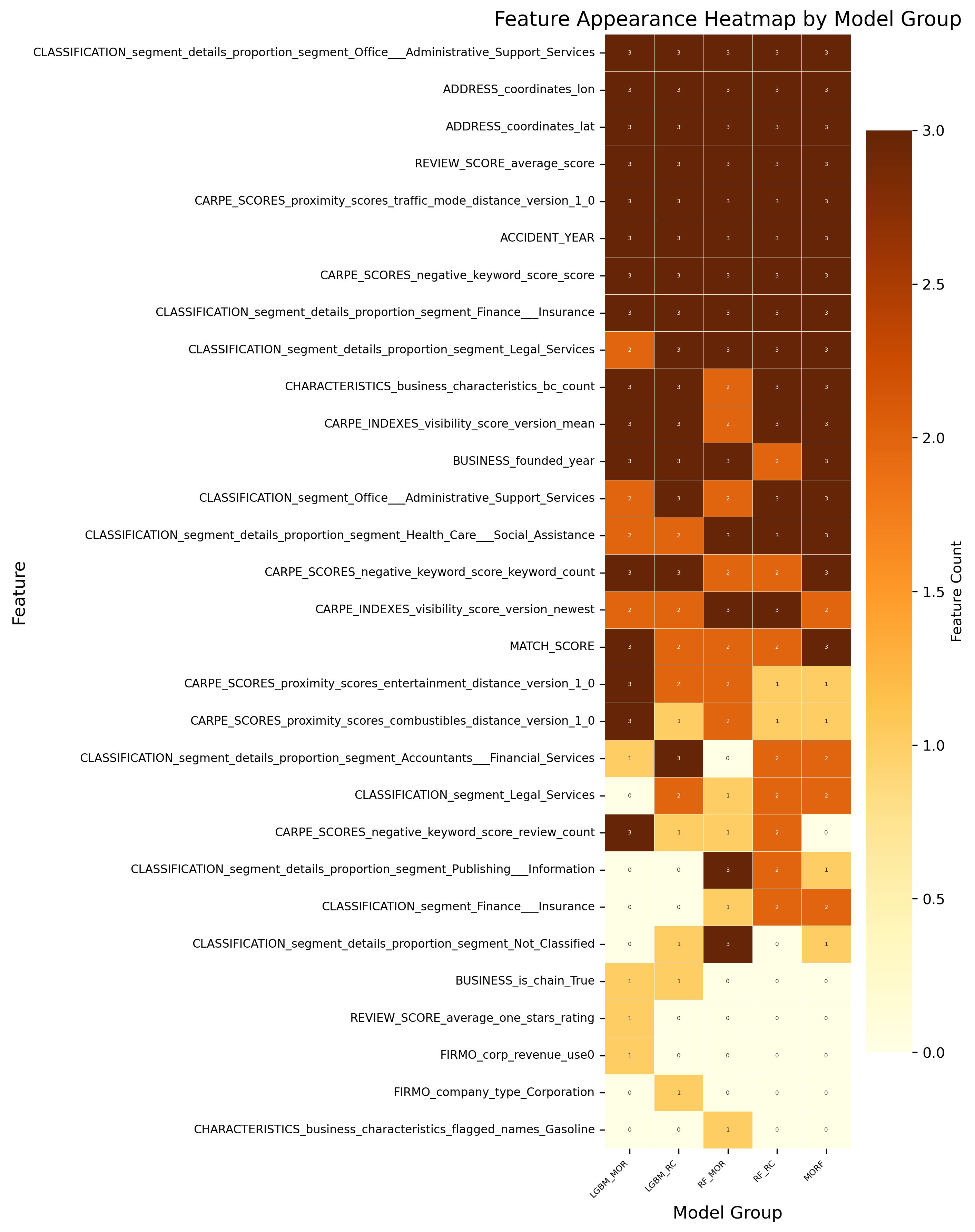}
        \caption{Count of top important feature appearance across various {regression} models.}
    \label{fig:feature_count_heatmap_reg}
\end{figure}

All analyses are conducted using the insurtech-enriched dataset (D2), which incorporates external entity-specific organizational features beyond the conventional cyber incident dataset. Figure~\ref{fig:importance_heatmap_cls} reports the log-transformed feature importance scores for the classification task. Each classification model produces multiple importance measures (e.g., impurity-based, permutation-based, SHAP-based). The features are ranked using their average importance across these measures, and the top 20 features under this ranking are visualized. The heatmap displays the importance values of these features across all individual measures. Figure~\ref{fig:feature_count_heatmap_cls} reports how consistently each feature is selected across the different importance measures within each classification model. A higher count reflects greater cross-measure stability, indicating that the feature is reliably identified as important regardless of the specific importance metric used. Analogous summaries for the regression models are presented in Figure~\ref{fig:importance_heatmap_reg} and Figure~\ref{fig:feature_count_heatmap_reg}. In both heatmaps, darker color intensity indicates higher importance or greater stability.


Building on the four heatmaps described above, we find that several features consistently stand out as influential, regardless of the underlying model architecture or the specific importance metric used. This consistency suggests that certain entity-specific organizational features play a systematically important role in capturing both incident occurrence and frequency. To provide a clearer and more accessible comparison of these recurring features, we extract the top 10 features based on the combined evidence from all four heatmaps. A consolidated summary of those results is presented in Appendix~\ref{app:notationE}, Table~\ref{tab:feature-importance-shortlist}. That table offers a direct, side-by-side overview of the dominant features, integrating information from all models and importance measures to highlight features that exhibit strong and persistent informative value across tasks.

Geographic indicators---a subset of \textit{Business Information} features in Table~\ref{tab:feature-summary-527}---consistently demonstrate predictive value across both tasks, though specific features differ by modeling objective. For occurrence modeling, address features (i.e., \textit{ADDRESS\_state\_east}, \textit{ADDRESS\_state\_west}, and \textit{ADDRESS\_street\_type\_St}) rank highly, suggesting that regional location and address structure influence the likelihood of incident occurrence, potentially reflecting regulatory variations or localized vulnerability. That is, some states implement more stringent security incident notification laws, increasing the likelihood of incidents being reported. Similarly, for frequency estimation, geographic coordinates (i.e., \textit{ADDRESS\_coordinates\_lon} and \textit{ADDRESS\_coordinates\_lat}) emerge as key features, capturing spatial heterogeneity in incident frequency patterns, likely related to geographic disparities in cyber exposure or reporting practices. 

Moreover, industry composition emerges as an important determinant for both occurrence and frequency estimation tasks. Sector-specific classification features (i.e., features beginning with \textit{CLASSIFICATION\_segment}) highlight that the industry sector composition affects not only the likelihood of incident occurrence but also its frequency. This result is consistent with findings in existing literature and current industry practices that regard industry classification as one of the most important rating factors. While broad sector-level classification features already rank highly across models, more granular measures of industry composition provide additional explanatory power. A key feature is the indicator for whether a firm is engaged in office and administrative support services (i.e., \textit{CLASSIFICATION\_segment\_details\_proportion\_segment\_office\_administrative\allowbreak\_support\_service}), which belongs to the \textit{Classification: Segment Proportion} category in Table~\ref{tab:feature-summary-527}. Its domain influence suggests that firms engaged more heavily in such sectors may be exposed to heightened cyber risk frequency, potentially due to their reliance on extensive administrative processes, routine handling of sensitive information, and higher volumes of digital interactions. This highlights that beyond sector membership, the inclusion of specific industry activities serves as a critical determinant of cyber incident risk.

Meanwhile, the number of business risk characteristics (i.e., \textit{CHARACTERISTICS\_business\_characteristics\_bc\_count}) is consistently prioritized in the occurrence and frequency modeling. This feature belongs to the broader \textit{Risk Characteristics} category mentioned in Table~\ref{tab:feature-summary-527} and its prioritization suggests that organizations with greater operational breadth or business diversity face a higher likelihood of experiencing cyber incidents, likely due to increased complexity or broader exposure surfaces.

While many influential features overlap between occurrence and frequency modeling, distinct task-specific patterns are also observed. For frequency estimation, features from the \textit{Proximity Score} category (i.e.,\textit{CARPE\_SCORES\_proximity\_scores\_traffic\_mode\_distance\_version\_1\_0}) emerge among the most influential features. Proximity scores capture how closely an organization is situated relative to surrounding activity or foot-traffic intensity. Their strong importance implies that firms located in more active or densely connected environments may experience higher incident frequency given their greater external exposure. This observation is also consistent with the high-importance geographic indicators we discussed earlier. Reputation-related features also show clear relevance in the frequency models. Features in the \textit{Index} category, such as \textit{REVIEW\_SCORE\_average\_score}, consistently contribute to explaining incident frequency, indicating that external customer-facing evaluations carry information about firms' underlying risk posture. In contrast, features in the \textit{Proximity} category play a more prominent role in the occurrence models. For example, \textit{CARPE\_SCORES\_negative\_keyword\_score\_keyword\_count} is repeatedly selected as an important feature for incident occurrence. Higher levels of negative online sentiment appear to be associated with an elevated probability of cyber incidents, suggesting that firms experiencing negative external signals may also face heightened operational exposure.

In addition to these entity-specific organizational features, the incident year (i.e., \textit{ACCIDENT\_YEAR}), which does not originate from the insurtech feature set but instead comes directly from the incident dataset, also emerges as one of the most influential features in both modeling tasks. Its strong predictive value likely reflects broader temporal dynamics, including shifts in the threat landscape, changes in reporting practices, and evolving regulatory requirements across years.

In summary, the consistent convergence across multiple importance measures indicates that many of the most influential features are concentrated in a few categories of entity-specific organizational features—\textit{Business Information} (e.g., geographic indicators), \textit{Classification} (e.g., segment proportions), and \textit{Proximity Score}. Within these categories, specific features repeatedly emerge as top-ranked features for both the occurrence and frequency of cyber incidents. This supports our central hypothesis that enriching sparse cyber incident data with entity-specific organizational features enables more granular and interpretable models for cyber risk.

\section{Discussion and conclusion} \label{sec:conclusion}

The modeling results suffice to answer the questions of interest raised in Section~\ref{sec:intro}. Entity-specific organizational features, in addition to industry classification and revenue, can offer substantial performance improvement in capturing the occurrence and frequency of cyber incidents. In particular, reputation-related features are found to play a pivotal role in affecting a firm's cyber risk, and this aspect has received limited attention and empirical validation in existing research. In addition, with the enriched dataset, the classification models demonstrate strong model performance in identifying the occurrences of cyber incidents. However, despite improvements driven by entity-specific organizational features, accurately predicting the frequencies of various types of incidents remains a challenging task. Lastly, we find little evidence for dependence among either occurrences or frequencies of different types of cyber incidents at the entity level. That is, once firm‑specific characteristics are taken into account, the occurrences and frequencies of different incident types are likely to be conditionally independent, and the occurrence or frequency of one incident type does not, in itself, provide predictive power for other types. The observed conditional independence is also supported by the almost negligible partial correlations of occurrences and frequencies across incident types after removing the effect of firm-specific characteristics (see Appendix \ref{app:dep}). This finding, however, does not address the unconditional dependence or independence among incident types.

To understand the performance gap between the frequency and occurrence models, we turn to the construction of our modeling dataset and the empirical characteristics of the cyber incident data. Two key factors drive the discrepancy. First, occurrence modeling is a binary classification task concerned only with identifying whether at least one incident occurs in a given firm-year. In contrast, frequency modeling requires predicting the number of incidents, which is significantly more complex. Cyber incident counts are highly skewed and dominated by zeros, leaving limited usable variation for estimating the conditional distribution of incident frequency. Second, the empirical distribution of cyber incidents is extremely imbalanced across incident types as well. As illustrated in Appendix~\ref {app:notation_dis}, Figure~\ref{fig:occurrence_dist}, the majority of occurrence records arise from \textit{Data Breach} and \textit{Privacy Violation}, whereas categories such as \textit{Extortion/Fraud}, \textit{IT Error}, and \textit{Other} appear only sporadically. A similar pattern emerges for incident frequency. As shown in Appendix~\ref {app:notation_dis}, Figure~\ref{fig:frequency_dist}, the total counts of incidents (i.e., aggregated frequency) exhibit a heavily right-skewed distribution, again dominated by \textit{Privacy Violation} and \textit{Data Breach}. This combination of modeling data create, imbalance, and heavy tails substantially reduces the effectiveness of frequency modeling, making it statistically more difficult than occurrence prediction.

Some limitations of this study are as follows and may motivate future research. For one, constrained by the availability of cyber loss data in the public domain, the study focuses only on the occurrence and frequency of cyber incidents. Whereas some empirical studies have examined the magnitude of cyber losses, none have conducted severity predictions using an extensive set of firm-specific features. Such severity models can potentially be established using proprietary datasets, such as the cyber loss data from Advisen and the claims data managed by cyber insurance providers. Moreover, beyond the insurtech-enhanced features explored in this study, there remains a significant opportunity for future enhancement in identifying and incorporating cyber risk factors. One promising direction is the integration of penetration test data, which leverages engineering domain expertise to replicate real-world attack environments. Such simulated scenarios could provide a richer understanding of system vulnerabilities and organizational response behaviors, offering valuable predictive signals that are difficult to capture through observational data alone. Incorporating such data could further strengthen the robustness and realism of cyber risk models.

Given the evolving and dynamic nature of cyber threats, it is increasingly urgent to develop a quantitative underwriting framework grounded in robust predictive analytics and sound risk management principles. By integrating high-frequency entity-specific data into underwriting models, insurers can improve risk differentiation, enhance pricing accuracy, and promote a more resilient cyber insurance market.

\section*{Acknowledgments}
Zhiyu Quan and Linfeng Zhang are supported by a 2024 Casualty Actuarial Society individual research grant. Any opinions, findings, conclusions, or recommendations expressed in this material are those of the authors and do not necessarily reflect the views of the Casualty Actuarial Society. Zhiyu Quan would like to thank the Gies College of Business at the University of Illinois Urbana-Champaign for its support of this work through the ORMIR Faculty Scholars Program. 

%\printbibliography

\section{References}

Biener, Christian, Martin Eling, and Jan Hendrik Wirfs. 2015. “Insurability of Cyber Risk: An Empirical Analysis.” \textit{The Geneva Papers on Risk and Insurance—Issues and Practice }40 (1): 131–58.

 Borchani, Hanen, Gherardo Varando, Concha Bielza, and Pedro Larranaga. 2015. “A Survey on Multi-Output Regression.” \textit{Wiley Interdisciplinary Reviews: Data Mining and Knowledge Discovery }5 (5): 216–33.

 Breiman, Leo. 2001. “Random Forests.” \textit{Machine Learning }45: 5–32.

 Breiman, Leo. 2002. \textit{Manual on Setting Up, Using, and Understanding Random Forests V3.1.} Statistics Department, University of California Berkeley.

 Cebula, James J., Mary E. Popeck, and Lisa R. Young. 2014. \textit{A Taxonomy of Operational Cyber Security Risks Version 2. }Carnegie-Mellon University. \href{https://apps.dtic.mil/sti/citations/tr/ADA609863}{https:}\href{https://apps.dtic.mil/sti/citations/tr/ADA609863}{//apps.dtic.mil/sti/citations/tr/ADA609863}.

 Chong, Wing Fung, Daniël Linders, Zhiyu Quan, and Linfeng Zhang. 2025. “Incident-Specific Cyber Insurance.” \textit{ASTIN Bulletin }55 (2): 395–425.

 Doe, Dissent. 2025. DataBreaches.net. \href{https://databreaches.net/}{https://databreaches.net/}. Accessed April 12. 

 Eling, Martin, and Kwangmin Jung. 2018. “Copula Approaches for Modeling Cross-Sectional Dependence of Data Breach Losses.” \textit{Insurance: Mathematics and Economics }82: 167–80.

 Eling, Martin, and Jan Wirfs. 2019. “What Are the Actual Costs of Cyber Risk Events?” \textit{European Journal of Operational Research }272 (3): 1109–19.

 Fisher, Aaron, Cynthia Rudin, and Francesca Dominici. 2019. “All Models Are Wrong, but Many Are Useful: Learning a Variable’s Importance by Studying an Entire Class of Prediction Models Simultaneously.” \textit{Journal of Machine Learning Research }20 (177): 1–81.

 Ganda, Dhatri, and Rachana Buch. 2018. “A Survey on Multi Label Classification.” \textit{Recent Trends in Programming Languages }5 (1): 19–23.

 Guo, Daya, Dejian Yang, Haowei Zhang, et al. 2025. “DeepSeek-R1: Incentivizing Reasoning Capability in LLMs via Reinforcement Learning.” Preprint, arXiv, \href{https://arxiv.org/abs/2501.12948}{https://arxiv.org/abs/2501.12948}.

 Harry, Charles, and Nancy Gallagher. 2018. “Classifying Cyber Events.” \textit{Journal of Information Warfare }17 (3): 17–31.

 Hinojosa Lee, Maria Cristina, Johan Braet, and Johan Springael. 2024. “Performance Metrics for Multilabel Emotion Classification: Comparing Micro, Macro, and Weighted F1-Scores.” \textit{Applied Sciences }14 (2): 9863.

 Hunt, Troy, Charlotte Hunt, and Stefán Jökull Sigurðarson. 2025. “Who’s Been Pwned.” Have I Been Pwned. \href{https://haveibeenpwned.com/PwnedWebsites}{https:}\href{https://haveibeenpwned.com/PwnedWebsites}{//haveibeenpwned.com/PwnedWebsites}. Accessed April 12, 2025.

 IBM. 2024. \textit{Cost of a Data Breach 2024. }\href{https://www.ibm.com/reports/data-breach}{https://www.ibm.com/reports/data-breach}. Accessed June 27, 2025.

 Ke, Guolin, Qi Meng, Thomas Finley, et al. 2017. “LightGBM: A Highly Efficient Gradient Boosting Decision Tree.” \textit{Advances in Neural Information Processing Systems }30.

 Kent, John T. 1983. “Information Gain and a General Measure of Correlation.” \textit{Biometrika }70 (1): 163–173.

 Kesan, Jay P., and Linfeng Zhang. 2020. “When Is a Cyber Incident Likely to Be Litigated and How Much Will It Cost? An Empirical Study.” \textit{Connecticut Insurance Law Journal }27 (2): 529–80.

 Kesan, Jay P., and Linfeng Zhang. 2021. “An Empirical Investigation of the Relationship Between Local Government Budgets, IT Expenditures, and Cyber Losses.” \textit{IEEE Transactions on Emerging Topics in Computing }9 (2): 582–96.

 Ko, Ryan, Elinor Tsen, and Sergeja Slapnicar. 2020. Dataset of Data Breaches and Ransomware Attacks over 15 Years from 2004. University of Queensland. \href{https://doi.org/10.14264/dfe5027}{https://doi.org/10.14264/dfe5027}.

 Lerman, Robert I., and Shlomo Yitzhaki. 1984. “A Note on the Calculation and Interpretation of the Gini Index.” \textit{Economics Letters }15 (3–4): 363–68.

 Lundberg, Scott M., Gabriel Erion, Hugh Chen, et al. 2020. “From Local Explanations to Global Understanding with Explainable AI for Trees.” \textit{Nature Machine Intelligence }2 (1): 56–67. 

 Lundberg, Scott M., and Su-In Lee. 2017. “A Unified Approach to Interpreting Model Predictions.” \textit{Advances in Neural Information Processing Systems }30.

 Nurse, Jason R. C., Louise Axon, Arnau Erola, Ioannis Agrafiotis, Michael Goldsmith, and Sadie Creese. 2020. “The Data That Drives Cyber Insurance: A Study into the Underwriting and Claims Processes.” In \textit{2020 International Conference on Cyber Situational Awareness, Data Analytics and Assessment (CyberSA).} IEEE.

 Palsson, Kjartan, Steinn Gudmundsson, and Sachin Shetty. 2020. “Analysis of the Impact of Cyber Events for Cyber Insurance.” \textit{The Geneva Papers on Risk and Insurance—Issues and Practice }45 (4): 564–79.

 Privacy Rights Clearinghouse. 2025. “Data Breach Chronology.” \href{https://privacyrights.org/data-breaches}{https://privacyrights.}\href{https://privacyrights.org/data-breaches}{org/data-breaches}. Accessed April 12, 2025.

 Quan, Zhiyu, Changyue Hu, Panyi Dong, and Emiliano A. Valdez . 2024. “Improving Business Insurance Loss Models by Leveraging InsurTech Innovation.” \textit{North American Actuarial Journal }29 (2): 247–74.

 Quan, Zhiyu, and Emiliano A. Valdez. 2018. “Predictive Analytics of Insurance Claims Using Multi-Variate Decision Trees.” \textit{Dependence Modeling }6 (1): 377–407.

 Read, Jesse, Bernhard Pfahringer, Geoff Holmes, and Eibe Frank. 2011. “Classifier Chains for Multi-Label Classification.” \textit{Machine Learning }85: 333–59.

 Rege, Aunshul, and Rachel Bleiman. 2023. “A Free and Community-Driven Critical Infrastructure Ransomware Dataset.” In \textit{Proceedings of the International Conference on Cybersecurity, Situational Awareness, and Social Media:} \textit{Cyber Science 2022; 20–21 June; Wales}\textit{,} edited by Cyril Onwubiko, Pierangelo Rosati, Anshul Rege, et al. Springer Nature.

 Romanosky, Sasha. 2016. “Examining the Costs and Causes of Cyber Incidents.” \textit{Journal of Cybersecurity }2 (2): 121–35.

 Romanosky, Sasha, Lillian Ablon, Andreas Kuehn, and Therese Jones. 2019. “Content Analysis of Cyber Insurance Policies: How Do Carriers Price Cyber Risk?” \textit{Journal of Cybersecurity }5 (1): tyz002. 

 Spyromitros-Xioufis, Eleftherios, Grigorios Tsoumakas, William Groves, and Ioannis Vlahavas. 2012. “Multi-Label Classification Methods for Multi-Target Regression.” Preprint, arXiv, arXiv:1211.6581.

 Tsohou, Aggeliki, Vasiliki Diamantopoulou, Stefanos Gritzalis, and Costas Lambrinoudakis. 2023. “Cyber Insurance: State of the Art, Trends, and Future Directions.” \textit{International Journal of Information Security }22 (3): 737–48.

 Tsoumakas, Grigorios, and Ioannis Katakis. 2008. “Multi-Label Classification: An Overview.” In \textit{Data Warehousing and Mining: Concepts, Methodologies, Tools, and Applications}. IGI Global.

 Verizon RISK Team. 2025. VCDB: VERIS Community Database. \href{https://github.com/vz-risk/VCDB}{https://github.com/vz-}\href{https://github.com/vz-risk/VCDB}{risk/VCDB}. Accessed April 12.

 Wheatley, Spencer, Annette Hofmann, and Didier Sornette. 2021. “Addressing Insurance of Data Breach Cyber Risks in the Catastrophe Framework.” \textit{The Geneva Papers on Risk and Insurance—Issues and Practice }46 (1): 53–78.

 Xu, Maochao, Kristin M. Schweitzer, Raymond M. Bateman, and Shouhuai Xu. 2018. “Modeling and Predicting Cyber Hacking Breaches.” \textit{IEEE Transactions on Information Forensics and Security }13 (11): 2856–71.

 Zängerle, Daniel, and Dirk Schiereck. 2023. “Modelling and Predicting Enterprise-Level Cyber Risks in the Context of Sparse Data Availability.” \textit{The Geneva Papers on Risk and Insurance—Issues and Practice }48 (2): 434–62.

 Zhang, Linfeng, Changyue Hu, and Zhiyu Quan. 2025. “NLP-Powered Repository and Search Engine for Academic Papers: A Case Study on Cyber Risk Literature with CyLit.” \textit{North American Actuarial Journal }29 (2): 390–421.

\clearpage
\clearpage

\begin{appendices}
\section{Notation}\label{app:notationA}
\begin{table}[!htbp]
\caption{Notation used in multilabel classification and multi-output regression.}
\centering
\label{tab:notation_classification}
\begin{tabular}{ll}
\toprule
\textbf{Symbol} & \textbf{Description} \\
\midrule
$\mathcal{D}$ & Dataset  \\
$\mathbf{x}_i \in \mathbb{R}^d$ & Feature vector for the $i$-th observation \\
$\mathbf{X} \in \mathbb{R}^{m \times d}$ & Full feature matrix (all $m$ observations) \\
$\mathbf{X}_{\text{train}} \in \mathbb{R}^{r \times d}$ & Training feature matrix \\
$\mathbf{X}_{\text{test}} \in \mathbb{R}^{s \times d}$ & Test feature matrix \\
$\mathbf{X}_{\text{val}} \in \mathbb{R}^{v \times d}$ & Validation feature matrix (split from $\mathbf{X}_{\text{train}}$ via cross-validation) \\
$\mathbf{y}_i \in \mathcal{Y} = \{0,1\}^q$ & Binary label vector for the $i$-th observation \\
$\mathbf{Y}_{\text{train}} \in \{0,1\}^{r \times q}$ & Training label matrix \\
$\mathbf{Y}_{\text{test}} \in \{0,1\}^{s \times q}$ & Test label matrix \\
$\mathbf{Y}_{\text{val}} \in \{0,1\}^{v \times q}$ & Corresponding label matrix for the validation fold \\
$f_{j,\theta}$ & Classifier/regressor for label $L_j$ trained with hyperparameter $\theta$ \\
$\mathbf{p}_j^{\text{val}} \in [0,1]^v$ & Predicted probabilities for label $j$ on validation set \\
$\mathcal{D}_j$ & Training dataset for label $L_j$ \\
$\mathcal{L} = \{L_1, \dots, L_j, \dots, L_q\}$ & Label set with $q$ classification output variables \\
$\hat{\mathbf{y}}_i \in \{0,1\}^q$ & Predicted label vector for the $i$-th observation \\
$\mathcal{T}$ & Thresholding strategy (FIXED or ADAPTIVE) \\
$\mathcal{B}$ & Base learner family (e.g., random forest) \\
$\Theta$ & Hyperparameter search space \\
$F $ & Evaluation metric \\
$m$ & Number of total observations in the dataset \\
$r$ & Number of training observations \\
$s$ & Number of test observations \\
$v$ & Number of validation observations (within training folds) \\
$d$ & Number of input features \\
$q$ & Number of labels (output dimensions) \\
$\mathbf{x}_i^* \in \mathbb{R}^d$ & Test observation for the $i$-th observation \\
$\hat{\mathbf{y}}_i^* \in \{0, 1\}^q$ & Final predicted binary label vector for the $i$-th observation \\
$\boldsymbol{\tau}^* = (\tau_1^*, \dots, \tau_q^*)$ & Vector of label-specific thresholds \\
$\hat{\mathbf{Y}}_{\text{test}} \in \{0,1\}^{s \times q}$ & Predicted binary label matrix for the test set\\
$K$ & Number of cross-validation folds \\
$\mathbf{z}_i \in \mathbb{R}^q$ & Continuous output vector for the $i$-th observation (e.g., incident counts) \\
\( \mathcal{O} = \{O_1, \dots, O_j, \dots, O_q\} \) & Label set with $q$ regression output variables \\
$\boldsymbol{\sigma} = (\sigma_1, \dots, \sigma_q)$ & Permutation of $\{1, \dots, q\}$ labels \\
$\Sigma$ & Total $q!$ permutations for the labels \\

\bottomrule
\end{tabular}

\end{table}

\clearpage

\section{Classification validation metrics}\label{app:notationB}
\begin{table}[!htbp]
\caption{Multilabel classification metrics with formulas and notations.}
\centering
\renewcommand{\arraystretch}{1.1}
\setlength{\tabcolsep}{5pt}
\small
\begin{tabular}{@{}>{\raggedright\arraybackslash}m{3.6cm} >{\raggedright\arraybackslash}m{3.5cm}>{\raggedright\arraybackslash}m{7.8cm}@{}}

\toprule
\textbf{Metric} & \textbf{Description} & \textbf{Formula} \\
\midrule

\textbf{Weighted-F1} & 
F1 weighted by label frequency & 
$\displaystyle \sum_{j=1}^{q} w_j \cdot \text{F1}_j$ , where $w_j = \frac{\text{Support}_j}{\sum_k \text{Support}_k}$, $\text{Support}_j = \text{TP}_j + \text{FN}_j$ \\
\midrule

\textbf{Macro-F1} & 
Unweighted mean of per-label F1 & 
$\displaystyle \frac{1}{q} \sum_{j=1}^{q} \text{F1}_j$ , where 
$\displaystyle \text{F1}_j = \frac{2 \cdot \text{TP}_j}{2 \cdot \text{TP}_j + \text{FP}_j + \text{FN}_j}$  \\
\midrule

\textbf{Micro-F1} & 
Global F1 across all labels & 
$\displaystyle \frac{2 \cdot \text{TP}}{2 \cdot \text{TP} + \text{FP} + \text{FN}}$  \\
\midrule

\textbf{Sample-F1} & 
Mean of sample-wise F1 scores & 
$\displaystyle \frac{1}{m} \sum_{i=1}^{m} \text{F1}_i$ , where  
$\displaystyle \text{F1}_i = \frac{2 \cdot \text{TP}_i}{2 \cdot \text{TP}_i + \text{FP}_i + \text{FN}_i}$ \\
\midrule

\textbf{Jaccard Index} & 
Intersection over union of label sets & 
$\displaystyle \frac{1}{m}\sum_{i=1}^m \frac{|\mathcal{S}_i \cap \hat{\mathcal{S}}_i|}{|\mathcal{S}_i \cup \hat{\mathcal{S}}_i|}$, where  $\mathcal{S}_i:=\{j: y_{ij}=1\},\; \hat{\mathcal{S}}_i:=\{j:\hat{y}_{ij}=1\}$ \\
\midrule

\textbf{Hamming Loss} & 
Proportion of incorrect labels & 
$\displaystyle \frac{1}{mq} \sum_{i=1}^{m} \sum_{j=1}^{q} \mathbb{I}(y_{ij} \ne \hat{y}_{ij})$ \\
\bottomrule
\end{tabular}

\label{tab:classification_metrics}
\end{table}

\section{Regression validation metrics}\label{app:notationC}
\begin{table}[!htbp]
\caption{Multi-output regression metrics with formulas and notations.}
\centering
\renewcommand{\arraystretch}{1.1}
\setlength{\tabcolsep}{5pt}
\small
\begin{tabular}{@{}>{\raggedright\arraybackslash}m{3.6cm} >{\raggedright\arraybackslash}m{3.5cm}>{\raggedright\arraybackslash}m{7.8cm}@{}}
\toprule
\textbf{Metric} & \textbf{Description} & \textbf{Formula} \\
\midrule

\textbf{average MSE (aMSE)} & Mean squared error across all outputs and observations &
$\text{aMSE} = \frac{1}{q} \sum_{j=1}^q \frac{1}{m} \sum_{i=1}^m (z_{ij} - \hat{z}_{ij})^2$  \\
\midrule

\textbf{average RMSE (aRMSE)} & Square root of aMSE per output &
$\text{aRMSE} = \frac{1}{q} \sum_{j=1}^q \sqrt{\frac{1}{m} \sum_{i=1}^m (z_{ij} - \hat{z}_{ij})^2}$ \\
\midrule

\textbf{average Correlation Coefficient (aCC)} & Mean Pearson correlation over outputs &
$\text{aCC} = \frac{1}{q} \sum_{j=1}^q \frac{\sum_{i=1}^m (z_{ij} - \bar{z}_j)(\hat{z}_{ij} - \bar{\hat{z}}_j)}{\sqrt{\sum_{i=1}^m (z_{ij} - \bar{z}_j)^2} \cdot \sqrt{\sum_{i=1}^m (\hat{z}_{ij} - \bar{\hat{z}}_j)^2}}$\\
\midrule

\textbf{average Relative RMSE (aRRMSE)} & RMSE normalized by true std per output &
$\text{aRRMSE} = \frac{1}{q} \sum_{j=1}^q \sqrt{\frac{\sum_{i=1}^m (z_{ij} - \hat{z}_{ij})^2}{\sum_{i=1}^m (z_{ij} - \bar{z}_j)^2}}$ \\
\midrule

\textbf{Global Euclidean Distance (EU\_DIST)} & Mean distance between predicted and true vectors &
$\frac{1}{m} \sum_{i=1}^{m} \left\| \mathbf{z}_i - \hat{\mathbf{z}}_i \right\|_2 = \frac{1}{m} \sum_{i=1}^{m} \sqrt{\sum_{j=1}^{q} (z_{ij} - \hat{z}_{ij})^2}$\\
\bottomrule
\end{tabular}

\label{tab:regression_metrics}
\end{table}
\clearpage

\section{Features from insurtech company}\label{app:notationD}

\begin{table}[!htbp]
\caption{Category-level summary of the 527 insurtech features used in analysis.}
\centering
\renewcommand{\arraystretch}{1.1}
\setlength{\tabcolsep}{4pt}
\small
\begin{tabular}{@{}>{\raggedright\arraybackslash}m{2.5cm}
                >{\raggedright\arraybackslash}m{3.5cm}
                >{\raggedright\arraybackslash}m{7.0cm}
                >{\raggedright\arraybackslash}m{2.2cm}@{}}
\toprule
\textbf{Category} & \textbf{Explanation} & \textbf{Examples} & \textbf{Number of Features} \\
\midrule

Business Information 
& Basic information describing business operations, including address, founding year, opening hours, menu, and other operational features 
& \textit{ADDRESS.coordinates.lat, BUSINESS.founded\_year, MENU.has\_menu\_True} 
& 114
\\ \hline

Firmographics 
& Features describing company size, workforce, ownership, revenue and sales, licensing count, and demographic features 
& \textit{FIRMO.company\_type\_Private, FIRMO.employee\_range\_11-50, FIRMO.revenue\_range\_1-4.9Mil, RANDOM\_STATS.is\_female\_owned\_False} 
& 50
\\ \hline

Classification 
& Categorical segmentation features describing the business segment and segment proportions 
& \textit{CLASSIFICATION.segment\_Retail, CLASSIFICATION.segment\_Automotive, CLASSIFICATION.segment\_details.proportion\_segment\_Manufacturing} 
& 48
\\ \hline

Risk Characteristics 
& Flags representing hazardous or risk-related business activities, services, physical features, and operational exposures 
& \textit{CHARACTERISTICS.business\_characteristics.flagged\_names.Alcohol, BUSINESS.professional\_misconducts\_count, URLS.has\_non\_corporate\_email\_True} 
& 211
\\ \hline

Index 
& Indicators derived from customer reviews that reflect various aspects of business quality, reputation, and service performance
& \textit{CARPE\_INDEXES.customer\_rating.score\_version\_mean, REVIEW\_SCORE.average\_score, REVIEW\_SUMMARY.Overall Business Rating} 
& 95
\\ \hline

Proximity Score 
& Environment-based indicators that measure risk arising from nearby businesses and surrounding neighborhood activity
& \textit{negative\_keyword\_score.score, proximity\_scores.combustibles.density\_v1.0, proximity\_scores.traffic\_mode.distance\_v1.0} 
& 9
\\

\bottomrule
\end{tabular}

\label{tab:feature-summary-527}
\end{table}

\clearpage

\section{Cross-model feature importance overview}\label{app:notationE}

\begin{table}[!htbp]
\caption{Summary of the top 10 important features in the classification and regression analyses.}
\centering
\renewcommand{\arraystretch}{1.1}
\setlength{\tabcolsep}{5pt}
\small
\begin{tabular}{@{}%
>{\raggedright\arraybackslash}m{7cm}%
>{\centering\arraybackslash}m{0.8cm}%
>{\centering\arraybackslash}m{0.8cm}%
>{\centering\arraybackslash}m{0.8cm}%
>{\centering\arraybackslash}m{0.8cm}%
>{\raggedright\arraybackslash}m{4cm}@{}
}
\toprule
\textbf{Feature} & \textbf{Cls-Imp} & \textbf{Cls-Const} & \textbf{Reg-Imp} & \textbf{Reg-Const} & \textbf{Feature Category} \\
\midrule

ACCIDENT\_YEAR & \checkmark & \checkmark & \checkmark & \checkmark 
& Cyber Incident Meta \\
\hline
ADDRESS\_coordinates\_lat &  &  & \checkmark & \checkmark 
& Business Information\\

ADDRESS\_coordinates\_lon &  &  & \checkmark & \checkmark 
& Business Information \\

ADDRESS\_state\_east &  & \checkmark &  &  
& Business Information\\

ADDRESS\_state\_west & \checkmark & \checkmark &  &  
& Business Information\\

ADDRESS\_street\_type\_St & \checkmark & \checkmark &  &  
& Business Information\\

BUSINESS\_founded\_year &  & \checkmark &  &  
& Business Information\\
\hline
CARPE\_SCORES\_negative\_keyword\_score\_keyword\_count 
& \checkmark & \checkmark &  &  
& Proximity Score\\

CARPE\_SCORES\_negative\_keyword\_score\_score 
&  &  &\checkmark  &  
& Proximity Score\\

CARPE\_SCORES\_proximity\_scores\_traffic\_mode\_distance\_version\_1\_0 
&  &  & \checkmark & \checkmark 
& Proximity Score\\
\hline
CHARACTERISTICS\_business\_characteristics\_bc\_count 
& \checkmark & \checkmark & \checkmark & \checkmark
& Risk Characteristics \\
\hline
CLASSIFICATION\_segment\_Finance\_\_Insurance 
&  &  &  &  \checkmark
& Classification \\

CLASSIFICATION\_segment\_Legal\_Services 
& \checkmark &  &  &  
& Classification\\

CLASSIFICATION\_segment\_Office\_\_Administrative\_Support\_Services 
& \checkmark & \checkmark &  & 
& Classification\\

CLASSIFICATION\_segment\_details\_proportion\_segment\_Finance\_\_Insurance 
& \checkmark & \checkmark & \checkmark & \checkmark 
& Classification\\

CLASSIFICATION\_segment\_details\_proportion\_segment\_Health\_Care\_\_Social\_Assistance 
& \checkmark & &  &  
& Classification\\

CLASSIFICATION\_segment\_details\_proportion\_segment\_Legal\_Services 
&  &  & \checkmark & \checkmark 
& Classification\\

CLASSIFICATION\_segment\_details\_proportion\_segment\_Office\_\_Administrative\_Support\_Services 
& \checkmark & \checkmark & \checkmark & \checkmark 
& Classification\\
\hline
REVIEW\_SCORE\_average\_score 
&  &  & \checkmark & \checkmark 
& Index\\
\bottomrule
\end{tabular}

\vspace{2mm}
\label{tab:feature-importance-shortlist}
\end{table}
\vspace{-3mm}
{\footnotesize
\textit{Notes:} 
(1) ACCIDENT\_YEAR is sourced from the cyber incident dataset; all other features originate from the insurtech dataset. 
(2) \textit{Cls-Imp} indicates whether a feature appears in the top 10 when features are ranked by their \emph{average} log-transformed importance across the different importance measures within each classification model, as summarized in Figure~\ref{fig:importance_heatmap_cls}. 
(3)\textit{Cls-Const} captures how consistently a feature is selected across the different importance measures within the same classification model; the top 10 is determined by the rank of counts, corresponding to Figure~\ref{fig:feature_count_heatmap_cls}. 
(4)\textit{Reg-Imp} and \textit{Reg-Const} are defined analogously for regression models, based on the summaries in Figures~\ref{fig:importance_heatmap_reg} and~\ref{fig:feature_count_heatmap_reg}.
}

\section{Unconditional and conditional dependence}\label{app:dep}

\subsection*{Dependence of occurrences}
The unconditional dependence of occurrences is computed using the occurrence labels in the training set from D2. The conditional dependence of occurrences is computed using the training residuals of {LGBM\_MOC} trained on D2. The association of residuals represents the dependence after the effect of firm-specific characteristics is removed. The phi coefficient is used to measure the degree of pairwise dependence of binary variables. 
\begin{table}[h]
\caption{Unconditional dependence and conditional dependence of occurrences across incident types.}
\centering
\resizebox{\textwidth}{!}{%
\begin{tabular}{@{}lrrrrr|rrrrr@{}}
\toprule
 & \multicolumn{5}{c|}{\begin{tabular}[c]{@{}c@{}}Unconditional Dependence of Occurrences \\ (Phi Coefficient)\end{tabular}} & \multicolumn{5}{c}{\begin{tabular}[c]{@{}c@{}}Conditional Dependence of Occurrences \\ (Phi Coefficient)\end{tabular}} \\ \midrule
 & \multicolumn{1}{c}{\begin{tabular}[c]{@{}c@{}}Privacy \\ Violation\end{tabular}} & \multicolumn{1}{c}{\begin{tabular}[c]{@{}c@{}}Data \\ Breach\end{tabular}} & \multicolumn{1}{c}{\begin{tabular}[c]{@{}c@{}}Extortion/\\ Fraud\end{tabular}} & \multicolumn{1}{c}{\begin{tabular}[c]{@{}c@{}}IT \\ Error\end{tabular}} & \multicolumn{1}{c|}{Other} & \multicolumn{1}{c}{\begin{tabular}[c]{@{}c@{}}Privacy \\ Violation\end{tabular}} & \multicolumn{1}{c}{\begin{tabular}[c]{@{}c@{}}Data \\ Breach\end{tabular}} & \multicolumn{1}{c}{\begin{tabular}[c]{@{}c@{}}Extortion/\\ Fraud\end{tabular}} & \multicolumn{1}{c}{\begin{tabular}[c]{@{}c@{}}IT \\ Error\end{tabular}} & \multicolumn{1}{c}{Other} \\
Privacy Violation & 1.00 & -0.75 & -0.24 & -0.16 & -0.07 & 1.00 & -0.02 & 0.02 & 0.02 & 0.02 \\
Data Breach & -0.75 & 1.00 & -0.13 & -0.13 & -0.04 & -0.02 & 1.00 & 0.03 & 0.05 & 0.03 \\
Extortion/Fraud & -0.24 & -0.13 & 1.00 & 0.02 & -0.02 & 0.02 & 0.03 & 1.00 & 0.03 & 0.00 \\
IT Error & -0.16 & -0.13 & 0.02 & 1.00 & -0.01 & 0.02 & 0.05 & 0.03 & 1.00 & -0.01 \\
Other & -0.07 & -0.04 & -0.02 & -0.01 & 1.00 & 0.02 & 0.03 & 0.00 & -0.01 & 1.00 \\ \bottomrule
\end{tabular}%
}

\label{tab:partial_corr_occ}
\end{table}

\subsection*{Dependence of frequencies}
The unconditional dependence of frequencies is computed using the frequency labels in the training set from D2. The conditional dependence of frequencies is computed using the training residuals of {LGBM\_MOR} trained on D2. The association of residuals represents the dependence after the effect of firm-specific characteristics is removed. Spearman's rho is used to measure the degree of pairwise dependence of numerical variables. 

\begin{table}[h]
\caption{Unconditional dependence and conditional dependence of frequencies across incident types.}
\centering
\resizebox{\textwidth}{!}{%
\begin{tabular}{@{}lrrrrr|rrrrr@{}}
\toprule
 & \multicolumn{5}{c|}{\begin{tabular}[c]{@{}c@{}}Unconditional Dependence of Frequencies \\ (Spearman's Rho)\end{tabular}} & \multicolumn{5}{c}{\begin{tabular}[c]{@{}c@{}}Conditional Dependence of Frequencies \\ (Spearman's Rho)\end{tabular}} \\ \midrule
 & \multicolumn{1}{c}{\begin{tabular}[c]{@{}c@{}}Privacy \\ Violation\end{tabular}} & \multicolumn{1}{c}{\begin{tabular}[c]{@{}c@{}}Data \\ Breach\end{tabular}} & \multicolumn{1}{c}{\begin{tabular}[c]{@{}c@{}}Extortion/\\ Fraud\end{tabular}} & \multicolumn{1}{c}{\begin{tabular}[c]{@{}c@{}}IT \\ Error\end{tabular}} & \multicolumn{1}{c|}{Other} & \multicolumn{1}{c}{\begin{tabular}[c]{@{}c@{}}Privacy \\ Violation\end{tabular}} & \multicolumn{1}{c}{\begin{tabular}[c]{@{}c@{}}Data \\ Breach\end{tabular}} & \multicolumn{1}{c}{\begin{tabular}[c]{@{}c@{}}Extortion/\\ Fraud\end{tabular}} & \multicolumn{1}{c}{\begin{tabular}[c]{@{}c@{}}IT \\ Error\end{tabular}} & \multicolumn{1}{c}{Other} \\
Privacy Violation & 1.00 & -0.72 & -0.23 & -0.15 & -0.07 & 1.00 & -0.60 & -0.09 & -0.04 & 0.01 \\
Data Breach & -0.72 & 1.00 & -0.13 & -0.13 & -0.04 & -0.60 & 1.00 & -0.14 & -0.12 & -0.03 \\
Extortion/Fraud & -0.23 & -0.13 & 1.00 & 0.02 & -0.02 & -0.09 & -0.14 & 1.00 & 0.08 & 0.00 \\
IT Error & -0.15 & -0.13 & 0.02 & 1.00 & -0.01 & -0.04 & -0.12 & 0.08 & 1.00 & 0.06 \\
Other & -0.07 & -0.04 & -0.02 & -0.01 & 1.00 & 0.01 & -0.03 & 0.00 & 0.06 & 1.00 \\ \bottomrule
\end{tabular}%
}

\label{tab:partial_corr_freq}
\end{table}

\clearpage
\section{Distribution of cyber incidents}\label{app:notation_dis}

\begin{figure}[!htbp]
    \centering
    \begin{subfigure}{0.95\textwidth}
        \centering
        \includegraphics[width=\textwidth]{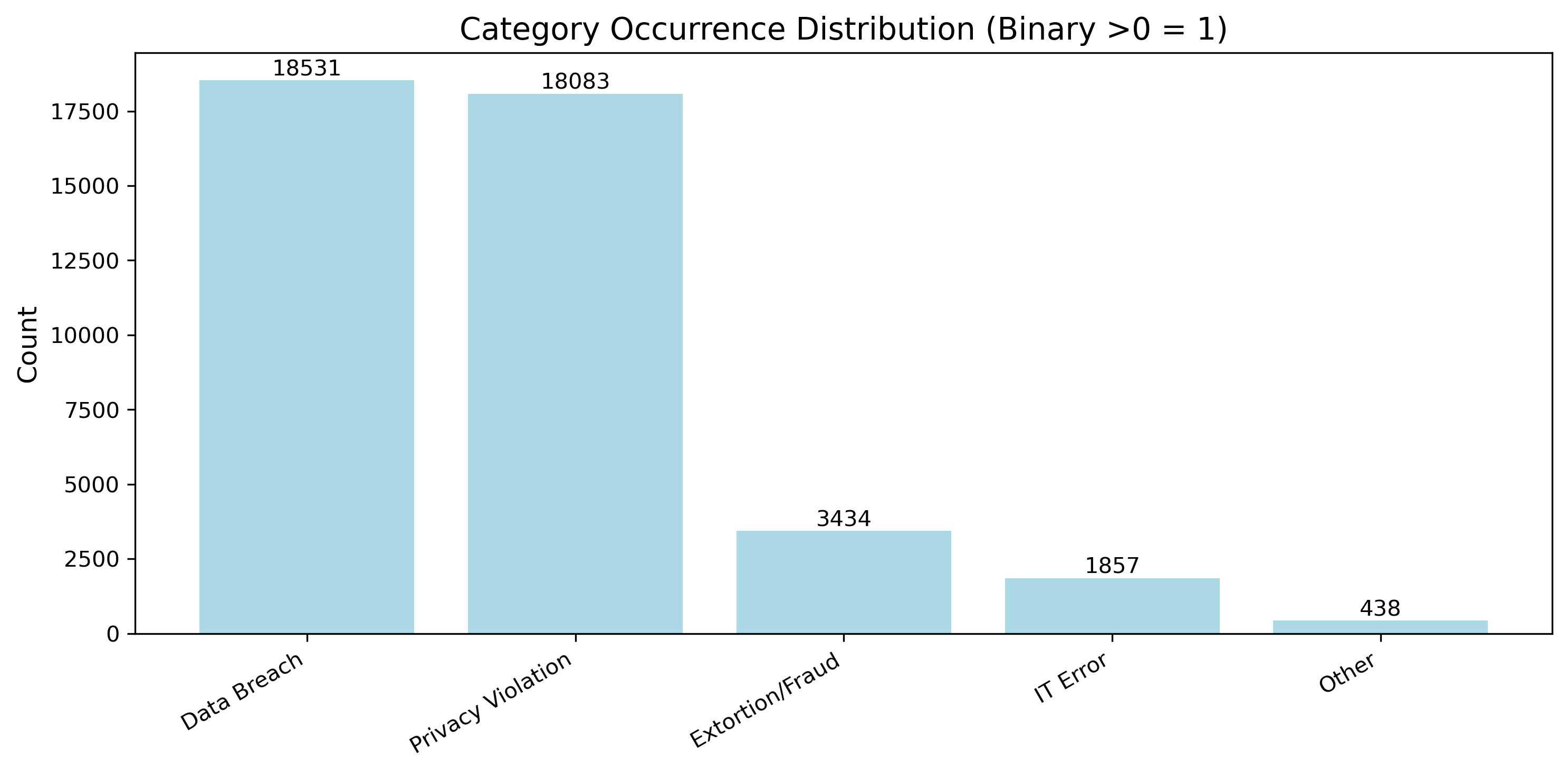}
        \caption{Distributions of cyber incident occurrence.}
        \label{fig:occurrence_dist}
    \end{subfigure}
    
    \vspace{0.4cm}
    
    \begin{subfigure}{0.95\textwidth}
        \centering
        \includegraphics[width=\textwidth]{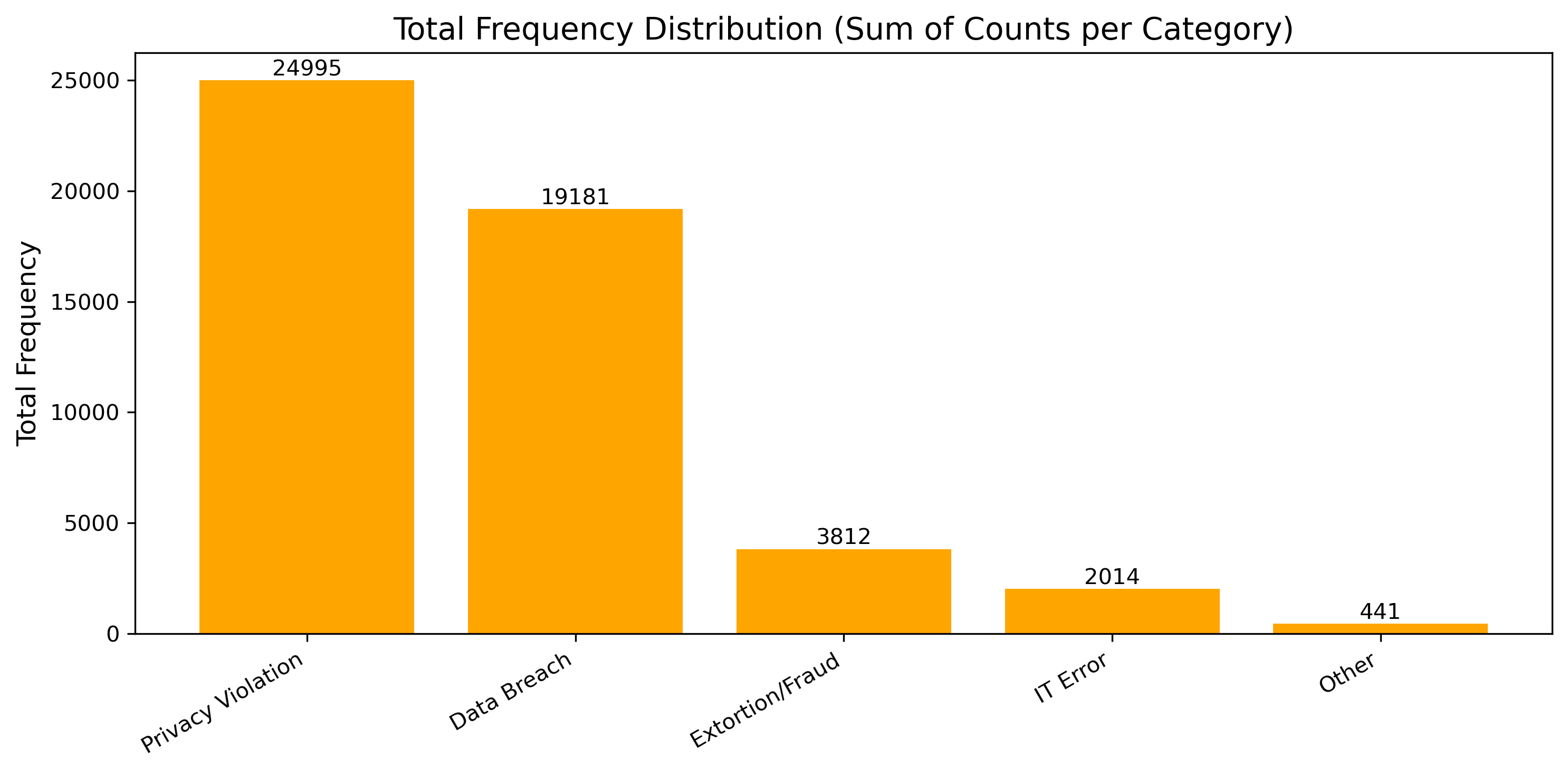}
        \caption{Distributions of cyber incident frequency.}
        \label{fig:frequency_dist}
    \end{subfigure}

    \caption{Distributions of cyber incident occurrence and frequency across categories.}
    \label{fig:incident_dist_combined}
\end{figure}
\end{appendices}
\end{document}